\documentclass[aip,
 pop,
 amsmath,amssymb,
 reprint,%
]{revtex4-1}

\usepackage{graphicx}
\usepackage{dcolumn}
\usepackage{bm}
\usepackage{xfrac}
\usepackage{epstopdf}
\usepackage{xcolor}

\begin{document}

\preprint{AIP/v0.0-POP}

\title{Resonant micro-instabilities at quasi-parallel collisionless shocks:\\cause or consequence of shock (re)formation\\
	}
\author{Vladimir Zekovi\'c}
 \email{vlada@matf.bg.ac.rs}
\affiliation{
	Department of Astronomy, Faculty of Mathematics, University of Belgrade\\Studentski trg 16, 11000 Belgrade, Serbia
}
\date{\today}

\begin{abstract}
A case of two interpenetrating, cold and quasi-neutral ion-electron plasmas is investigated with the multi-fluid approach. We consider that one plasma flows quasi-parallel to the lines of a background magnetic field embedded in another static plasma. If the flow turns super-Alfv\'enic, we show that parallel R/L-modes and perpendicular X/O-modes become unstable and grow in amplitude. Within the linear theory, we find that the growth rate curve of an unstable mode has a maximum at some wavenumber specific to each mode. If we consider a shock-like plasma configuration, we find that the fastest growing mode is the resonant one (with $k \sim r_{gi}^{-1}$) which strongly interacts with ions. In Particle-In-Cell (PIC) simulations, we observe that a resonant wave with the same properties is excited during the early phases of shock formation. Once the wave becomes non-linear, it efficiently scatters ions and triggers the initial shock formation. This implies that the typical compression ratio of $\sim 4$ could naturally arise as a consequence of a highly resonant micro-physical process. We model the interaction of ions reflected from the reforming shock barrier in a weak-beam case, and we show that the upstream wave now matches the instability we expect from the equations. By using PIC simulations, we explain how the strong-beam resonant instability triggers shock formation in the non-linear stage, and how the weak-beam instability reforms and transmits the shock afterwards. The micro-instabilities that we study here could largely contribute to shock triggering as well as to the reformation and transmission of the shock itself.
\end{abstract}

\pacs{52.35.Py, 52.35.Bj, 52.35.Tc, 52.30.-q, 52.65.Rr}

\maketitle

\section{\label{sec:intro}Introduction}

Although studied theoretically for many years and observed in our space surroundings, collisionless shocks still show a variety of physical processes that reveal complex physics behind them. Some basic properties of shocks in general, including non-linear steepening, are explained by MHD theory and simulations. Nevertheless, such models could not account for particle acceleration to cosmic-ray (CR) energies and the formation of non-thermal particle spectra. When the first-order Fermi acceleration mechanism, known as diffusive shock acceleration (DSA), was initially introduced~\cite{Krymsky,Axford,Bell781,BlandOstr,Drury}, it was very successful in explaining these important features of shocks. However, the picture of how particles gain high enough kinetic energy so that they can be injected into DSA was not completely clear. Later, particle-in-cell (PIC) simulations showed how this mechanism may work in the case of quasi-parallel shocks. From PIC and kinetic hybrid simulations~\cite{sda,nrelshocks}, we know that particles are at first specularly reflected and pre-accelerated by an upstream motional electric field ${\bf E}_0 = - {\bf v}_0 \times {\bf B}_0$ (where $v_0$ is the velocity of the flow and $B_0$ is the background magnetic field). While particles drift along the shock surface, they gyrate around the mean magnetic field, and they gain energy by $E_0$ in every cycle. This process, known as shock drift acceleration (SDA)~\cite{sdaf}, is a dominant acceleration mechanism at quasi-perpendicular shocks. At quasi-parallel shocks, particles are energized up to a few times $E_\text{sh} = m v_\text{sh}^2 / 2$ by the same mechanism. Particles, thus, become supra-thermal, and those that return upstream, contribute to the formation of a return current. This current excites the instability~\cite{NR,amatoblasi} in the upstream, which further accelerates particles to energies much higher than $\sim E_\text{sh}$, through a DSA-like process.

To understand how a return current of ions is formed, we need to consider the fine structure of a collisionless shock. The emerging possibility that the steepening of a wave, which is essential to the formation of a collisional shock, is not a required condition in the case of a collisionless shock, opens a question about the processes that trigger and shape the shock structure in an unstable and turbulent plasma. These processes are highly non-linear, but the theory of corresponding micro-physics is mostly linear and gives a more qualitative than quantitative explanation. Kinetic simulations can provide us with detailed insight into how this rather complex structure forms and how it changes over time. The transition region at quasi-parallel shocks is shown to be much wider than at quasi-perpendicular ones~\cite{nrelshocks,sda} (which are more similar to MHD shocks). This shock interface shows a variety of different types of instabilities and turbulences that can grow, then saturate or disrupt, or even mutually overlap. The type of these changes and their dynamics depend on the strength and level of magnetization of the shock itself. Recent papers~\cite{QPARCSHKs,TMPHCSHKW} review the efforts made to explain such systems and how numerical simulations, observations, and theory agree.

In kinetic simulations of quasi-parallel shocks, filamentation instability can be induced by CRs~\cite{filament}, which first excite the transverse modes in the upstream. The shock itself could result from the non-linear steepening of these modes~\cite{sda}, which can occur quasi-periodically after every few gyrations; this is a phenomenon also known as the shock reformation process~\cite{onshckreform,locshckreform}. In case of a quasi-perpendicular shock, an incoming flow is reflected off the compressed magnetic field. At weakly magnetized or unmagnetized shocks~\cite{unmagnetized}, the Weibel instability forms current and density filaments in the overlapping region. These filaments merge (through the second Weibel instability) and disintegrate, thus forming the shock transition.

These different types of shock interfaces, together with a reforming shock barrier, reveal the complexity of the system. This was the motive for investigating whether there is some other type of wave-plasma interaction that can quickly put, from ground state, the two interpenetrating collisionless plasmas into a shock configuration. On the microscopic level of this interaction, particles should be effectively scattered, and their motion randomized. To make the closure and fulfill the pre-acceleration conditions, we require this interaction to be able to form the initial return current. We also require that the same type of interaction later starts the reformation process.

We assume that at the very beginning of the shock formation, a finely tuned resonant wave-particle interaction occurs at the interface of a moving and steady plasma. We observe this in the PIC simulations presented here. As the interface propagates with the plasmas, this interaction can contribute to the thermalization by scattering the particles resonantly. The scattering should occur on the scales of the order of a particle gyroradius. At such scales, most of the particles will become trapped by the wave. If there is no turbulence, the resonance acts to prevent the particles from leaking even from the region a few gyroradii away from the interface. Still, some of the particles can be reflected at the edge of the interface with a velocity fast enough to escape the wave. This scenario is therefore restricted to super-critical shocks with magnetic inclination $\lesssim 45^\circ$.

The properties of such an idealized resonant interaction were examined in the previous work~\cite{rpbsw}. Therein, we assumed that the growth of circularly polarized EM waves is seeded and governed at the interface, where the bulk plasma flow interpenetrates the steady plasma. Using EM N-body simulations, we studied the motion of test particles in the field of circularly polarized ion-cyclotron frequency ($\sim \omega_{ci}$) waves. We showed that the scattering of test ions by this wave is the strongest, if the wavelength of such a mode is resonant ($\lambda \sim \pi r_{gi}$) with respect to the longitudinal gyroradius ($r_{gi} = v_\parallel / \omega_{ci} = m v_\parallel / q B_0$) of the ions inflowing with the velocity component $v_\parallel$ (parallel to $B_0$). Not only are the particle velocity vectors then randomized along the direction of the flow, but there is also a population of particles that are resonantly backscattered (reflected) from the wave. We found that the initial return current of reflected ions can be quite strong ($\gtrsim 10 \%$ of the ions impinging on the wave are reflected). We also showed that if the wavelength is resonant, the flow disrupts even when the wave amplitude is as low as $\approx B_0$ in the non-linear regime. For shorter or longer wavelengths, the amplitude of a few times $B_0$ or greater is required for a monochromatic wave to efficiently scatter and disrupt the flow.

In this paper, we use these empirical results as tracers to search for the equivalent interaction in the linear plasma theory, which naturally favors the growth of resonant modes. We also use these results to calibrate the derived equations so as to analytically show how such resonant instability grows and triggers the formation of a shock without the involvement of any other turbulent mechanisms. We here associate the resonant wave that appears in PIC simulations at the plasma interface with the strong-beam instability. The upstream wave we model as a weak-beam instability. Both instabilities are induced by the same type of beam-plasma interaction, and they differ only in the strength of the beams that excite them. The first refers to shock triggering, and the second refers to shock reformation. By using PIC simulations, we examine the behavior of these waves in the non-linear regime.

We organized the paper as follows: the equations governing the propagation of plasma waves and stability criterion are derived in Sec.~\ref{sec:resalfinst}; the theoretical results obtained here are applied to shock waves and compared to the results of numerical simulations in Sec.~\ref{sec:pic}; and finally, the possible role of the resonant instabilities in triggering the collisionless shock formation along with their role in governing the shock reformation process is discussed in Sec.~\ref{sec:results}.

\section{\label{sec:resalfinst}Cyclotron micro-instability}

We start from the fundamental case of two cold and quasi-neutral ion-electron plasmas that collide in the presence of a background magnetic field. We consider that one plasma (of a finite extent) flows quasi-parallel to the magnetic field lines with the velocity $v_0$, through another static plasma. In the case of wave frequencies much lower than $\omega_{ci}$, recent studies~\cite{rusmhd,awcfdm} show that linearly polarized Alfv\'en waves can grow and become coupled with the flow-driven ($\omega = k v_0$) instability. Such instabilities can draw the energy for their growth from the macroscopic kinetic energy pool of interpenetrating plasmas and the background magnetic field. The equations show that in the frame of a flowing plasma, the instability which starts to grow at the edge of the interface, advects further away with the stream. Thus, we have the plasma flowing into the interface with the velocity $-v_0$, and the instability advecting away from the interface at a velocity of $- v_0 / (1 + \eta)$. Therefore, if in the non-linear regime the wave is capable of triggering the scattering of the particles, it can easily provide a plasma configuration which by its dynamics resembles a shock wave. However, the instability growth rate has linear dependence with $k$, and the solutions are restricted only to low-frequency waves. None of the resonant $k$ is favored within this theory, which thus fails to provide the solution for the predecessor wave in the linear regime.

To include frequencies up to the ion-cyclotron $\omega_{ci}$ and electron-cyclotron $\omega_{ce}$, each of the plasma species must be treated separately. Moreover, if we confine to the space plasmas that can be considered as collisionless, then the Vlasov equation (or the fluid equations derived from it) should be used. From the linear theory and simulations, we know that various electrostatic and EM micro-instabilities~\cite{microinst,beaminst} can grow within a configuration where the two plasmas collide with a relative drift velocity $v_0$ parallel or anti-parallel to an average (or uniform) background magnetic field. The most important are the circularly polarized EM two-component modes, such as ion/ion, electron/electron and ion/electron modes, which have already been studied in literature~\cite{microinst}. The ion/ion resonant mode~\cite{emibinst} is the most interesting among them, since it is immune to the small changes of the plasma parameters because of its large wavelength. Its growth rate has a maximum near the wavenumbers at which the real frequency part satisfies a cyclotron resonance condition $\omega_r \simeq {\bf v}_0 \cdot {\bf k} \pm \omega_{ci}$. It has large application to solar wind and terrestrial bow shock~\cite{emiiinst}, where it appears as a right-hand resonant instability initiated by a return current of ions in the upstream region. However, these modes are associated with relatively weak particle beams.

We consider here the modes excited by a strong plasma beam, whose density is comparable to target plasma density. The beam and target plasmas consist of both species, ions and electrons. Note that in the case considered here, the resonance we refer to is not given by the cyclotron resonant condition $\omega_r \simeq {\bf v}_0 \cdot {\bf k} \pm \omega_{ci}$ as in~\cite{microinst,emibinst}. Since the beam is quite strong, all terms in the dispersion relation become significant, so the resonance does not occur in the same way. In this case, the resonance is given by the condition that the wavelength of the resonant mode $\lambda_\text{res}$ is of the order of $r_{gi}$. We also assume that the relative drift velocity of the two plasmas is much larger than the average thermal velocity of the plasma components ($v_0 \gg v_{T}$). The equations can then be written in the cold plasma limit (closed by the equation of state $p=0$, where $p$ corresponds to the hydrodynamic pressure).

\subsection{\label{sec:equations}Equations of cold interpenetrating plasmas}

Fluid equations of cold plasma consisting of ions and electrons are given by~\cite{intro2plasma}:

\begin{eqnarray}
	m_i \left( \frac{\partial}{\partial t} + {\bf V}_i \cdot \nabla \right) {\bf V}_i = q_i \left( {\bf E} + {\bf V}_i \times {\bf B} \right),
	\label{eq:nli}
	\\
	m_e \left( \frac{\partial}{\partial t} + {\bf V}_e \cdot \nabla \right) {\bf V}_e = q_e \left( {\bf E} + {\bf V}_e \times {\bf B} \right),
	\label{eq:nle}
\end{eqnarray}

\noindent where ${\bf E}$ and ${\bf B}$ are electric and magnetic field components, $\rho_{i,e} = n_{i,e} m_{i,e}~$ is the density of ions (electrons), and $q_{i,e}$ their charge. For the flowing plasma, the velocity of each species has the same constant component ${\bf v}_0$ added to its variable part ${\bf v}_{i,e}$, so that the total fluid velocity is ${\bf V}_{i,e} = {\bf v}_0 + {\bf v}_{i,e}$. For each variable, we consider perturbations of the form $f({\bf r},t) = f({\bf k},\omega) \cdot e^{i({\bf k r} - \omega t)}$.

The dispersion matrix that constitutes a system of linearized equations (as derived in Sec.~\ref{sec:appequations} in the Appendix) is

\begin{eqnarray}
	\mathcal{D} &&=
	\begin{pmatrix}
		P - n^2 \sin^2 \theta & 0 & n^2 \sin\theta \cos\theta \\
		0 & S - n^2 & -i D \\
		n^2 \sin\theta \cos\theta & i D & S - n^2 \cos^2 \theta\\
	\end{pmatrix} + \nonumber
	\\
	&&+ \frac{i}{\epsilon_0 \omega} \sigma_f,
	\label{eq:gdm}
\end{eqnarray}

\noindent where $\theta$ is the angle between the wavevector $\bf k$ and the background magnetic field ${\bf B}_0$; and $n$ is the refractive index. The components $P$, $S$, $D$ and matrix $\sigma_f$ are given by Eqs.~\ref{eq:polmkp}, \ref{eq:polmks}, \ref{eq:polmkd} and \ref{eq:polms}, respectively. Dispersion relation is found by equating the determinant of the dispersion matrix to zero, $|\mathcal{D}| = 0$.

Two specific cases are considered in the following analysis -- solutions to the dispersion relation for wavevector $\bf{k}$ parallel (Sec. \ref{sec:parallel}) and normal (Sec.~\ref{sec:normal}) to the background magnetic field ${\bf B}_0$.

\subsection{\label{sec:parallel}Wavevector parallel to the magnetic field}

Without loss of generality, the vector of the background magnetic field is taken to be in the direction of the $x$-axis (${\bf B}_0 = B_0 {\bf \hat{x}}$). In case of $\theta = 0$, the unit wavevector is ${\bf \hat{k}} = {\bf \hat{x}}$, and the dispersion matrix (\ref{eq:gdm}) reduces to

\begin{eqnarray}
	\mathcal{D} &&=
	\begin{pmatrix}
		P & 0 & 0 \\
		0 & S - n^2 & -i D \\
		0 & i D & S - n^2\\
	\end{pmatrix} - \nonumber
	\\
	&&- \sum_{\alpha = i,e} \frac{\omega_{p\alpha f}^2}{\omega^2} \frac{k_x v_0^x}{\xi \omega} \cdot
	\begin{pmatrix}
			2 + \dfrac{k_x v_0^x}{\xi \omega} & 0 & \ \ 0 \\
			0 & 0 & \ \ 0 \\
			0 & 0 & \ \ 0 \\
	\end{pmatrix},
	\label{eq:dmp}
\end{eqnarray}

\noindent where $\omega_{p\alpha f}$ is the plasma frequency of the species $\alpha$ in the flowing plasma, and $\xi = 1 - {\bf v}_0\cdot{\bf k} / \omega$ is the modification parameter of the frequency due to the flow velocity. Dispersion relation is then found as

\begin{eqnarray}
	\left[ P - \sum_{\alpha = i,e} \frac{\omega_{p\alpha f}^2}{\omega^2} \frac{k_x v_0^x}{\xi \omega} \cdot \left( 2 + \dfrac{k_x v_0^x}{\xi \omega} \right) \right] \cdot \label{eq:drp}
	\\
	\cdot \left[ (S-n^2)^2-D^2 \right] = 0, \nonumber
\end{eqnarray}

\noindent whose roots are assorted according to the associated eigenvectors of the electric field.

The first root branch of this equation corresponds to the dispersion relation of a longitudinal electrostatic oscillation advected by the flow. After substituting $P$ from Eq.~(\ref{eq:polmkp}) and by making the use of the equivalence $k_x v_0^x \equiv {\bf k}\cdot{\bf v}_0$, which is applicable in this parallel case, the first root becomes

\begin{equation}
	1 - \sum_{\alpha = i,e} \frac{\omega_{p\alpha s}^2}{\omega^2} - \sum_{\alpha = i,e} \frac{\eta~\omega_{p\alpha s}^2}{(\omega - {\bf k}\cdot{\bf v}_0)^2} = 0,
	\label{eq:drplong}
\end{equation}

\noindent where the density ratio of the flowing to the stationary plasma is given by $\eta = \omega_{pif}^2 / \omega_{pis}^2 = \omega_{pef}^2 / \omega_{pes}^2$. This is the standard cold plasma dispersion relation for electrostatic waves, with the inclusion of the velocity drift between the different plasma species~\cite{el2stream,el2stream2}. Depending on the wave frequency, a longitudinal electrostatic wave can be coupled with and advected by the flow. This mode can become unstable when the electrostatic wave and the streaming plasma are close to being in resonance $\omega \to {\bf k}\cdot{\bf v}_0$. Energy is thus transferred from the electrons to the wave, leading to the exponential growth of the wave amplitude. As shown in~\cite{el2stream3}, the waves saturate by trapping the electrons, and then collapse, making the electrons become rapidly thermalized.

The second root branch of Eq.~(\ref{eq:drp}) is given by

\begin{eqnarray}
	n^2 &&= S \pm D = R (L), \nonumber
	\\
	R (L) &&= 1 - \nonumber
	\\
	&&- \sum_{\alpha = i,e} \omega_{p\alpha s}^2 \left[ \frac{\eta~ \xi^2}{\xi \omega (\xi \omega \pm \omega_{c\alpha})} + \frac{1}{\omega (\omega \pm \omega_{c\alpha})} \right],
	\label{eq:drpnormrl}
\end{eqnarray}

\noindent which is essentially the same as in the case~\cite{microinst} of a two-species, three-component cold plasma consisting of a tenuous beam that interpenetrates a dense core and the third component of the other species.

These right (R) and left (L) circularly polarized waves given by Eq.~(\ref{eq:drpnormrl}) can both turn into instabilities. For comparable densities of the flowing and stationary plasmas, two types of instabilities can occur. One of them is resonant and it has $\lambda \sim \pi r_{gi}$. The other type is sub-resonant in the sense that its wavelength is of the order of the longitudinal gyroradius of streaming ions ($\lambda \sim r_{gi}$). These two modes are very similar to EM ion/ion modes, as given in~\cite{microinst}. We here solve the equations of each of the modes analytically (see Sec.~\ref{sec:appparallelinst} in the Appendix), which is only possible when considering some specific range of frequencies. Real ($\omega_r$) and imaginary ($\gamma$) parts of the frequency $\omega = \omega_r + i \gamma$ are thus derived as solutions to the quartic equation.

In the range of ion-cyclotron frequencies, condition $\xi \omega \sim \omega_{ci} \ll \omega_{ce}$ implies the interaction between ions of the flowing and stationary plasmas. Eq.~(\ref{eq:drpnormrl}) is therefore approximated by

\begin{equation}
	R (L) \approx 1 + \omega_{pis}^2 \left[ \frac{\eta~ \xi^2}{\omega_{ci} (\omega_{ci} \pm \xi \omega)} + \frac{1}{\omega_{ci} (\omega_{ci} \pm \omega)} \right].
	\label{eq:rlinstii3}
\end{equation}

For frequencies much lower than $\omega_{ci}$, two of the four roots of this relation merge to form the solutions which are the same as in Ref.~\cite{awcfdm}. Eq.~(\ref{eq:rlinstii3}) therefore represents the generalization of the case where at low frequencies an Alfv\'en wave couples with the flow driven ($\omega = k v_0$) mode~\cite{awcfdm}.

\begin{figure}[h!]
	\includegraphics[width=\linewidth]{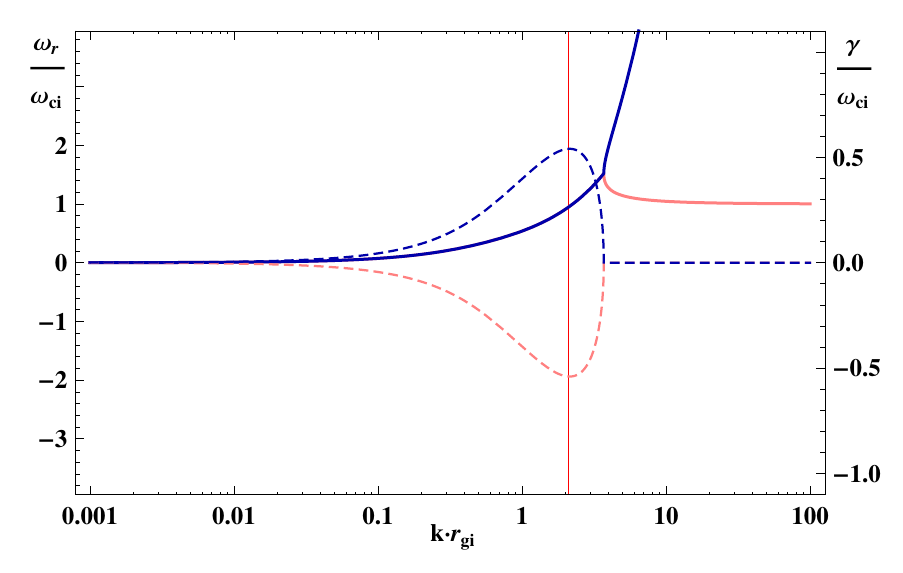}
	\caption{\label{fig:Liwy} The L-wave real frequency and growth rate dependences of the wavenumber for plasmas with the density ratio $\eta = 3$. The real part ($\omega_r$) of each of the roots is plotted as a continuous line and the imaginary part ($\gamma$) as a dashed line; both expressed in relative units of $\omega_{ci}$, while the wavenumber is given in units of $r_{gi}^{-1}$. The curves of $\omega_1$ and $\omega_2$ are given in dark blue and red color, respectively. The position of the maximum is at the resonant wavenumber $k_{L_\text{max}} = 2 \pi / \lambda_\text{res} \approx 2/r_{gi}$.}
\end{figure}

These two roots, given by the quartic pairs $\omega_{3,4}$ or $\omega_{1,2}$, are thus both assigned to the R or L-mode, respectively, depending on whether ``$+$'' or ``$-$'' sign is chosen in Eq.~(\ref{eq:wr4i}) in the Appendix. Real and imaginary parts of the complex frequency $\omega_{1...4} = \omega_r + i \gamma$ correspond to the real frequency and growth rate of the wave, respectively. The two roots of the L-mode equation are plotted in Fig.~\ref{fig:Liwy}. One of them grows with $\gamma \approx 0.55~\omega_{ci}$, while the other one decreases at the same rate.

Earlier, we determined by test particle simulations~\cite{rpbsw} that a beam of ions interacts with the L-wave only in a very narrow band of wavelengths around $\lambda_\text{res} \approx \pi r_{gi}$, where $r_{gi} = m_i v_0 / q_i B_0$ and $v_0$ is the velocity of the beam. From Fig.~\ref{fig:Liwy}, we observe that both roots have a maximum at some wavenumber $k_{L_\text{max}}$. If we impose the condition that the maximum must be located exactly at $k_{L_\text{max}} = 2 \pi / \lambda_\text{res} = 2 / r_{gi}$ in the equations, they then give the density ratio of the two plasmas. Therefore, when the wave is resonant, this ratio is $\eta \approx 3$. Or to put it another way, if the flowing plasma density is three times the density of the stationary plasma, the L-mode (that emerges from the wave-plasma interaction) is then resonant. Near the beginning of the overlapping region, where this resonant wave grows, there is an overall density ratio of $\eta + 1 \approx 4$. The linear equations thus show that the resonant wavenumber $k_{L_\text{max}}$ is related to the shock-like density ratio at the interface of the two interpenetrating plasmas. Although this connection may be a coincidence, it still implies that the resonant or sub-resonant modes should grow the fastest.

\begin{figure}[h!]
	\includegraphics[width=\linewidth]{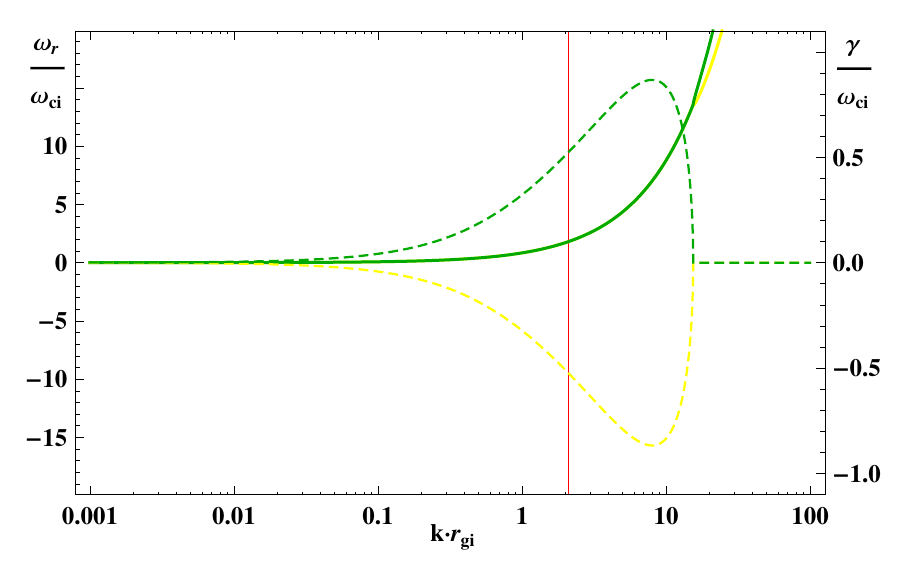}
	\caption{\label{fig:Riwy} Same as in Fig.~\ref{fig:Liwy}, only for the R-wave. The curves of $\omega_3$ and $\omega_4$ are given in green and yellow color, respectively. The growth of the wave is seeded at the sub-resonant wavelengths $k_{R_\text{max}} = 2 \pi / \lambda_\text{sub} \approx 8/r_{gi} \approx 4 k_{L_\text{max}}$. The position of the resonance $k_{L_\text{max}}$ is marked by a vertical red line.}
\end{figure}

There is a pair of solutions to the R-mode that also has the non-zero complex part of the frequency, which is shown in Fig.~\ref{fig:Riwy}. It corresponds to a sub-resonant (with the ions gyroradii) wave-mode. For comparable densities of the two colliding plasmas, the wavenumber of the maximum growth of the sub-resonant wave $k_{R_\text{max}}$ is roughly proportional to the maximum of the resonant mode as $k_{R_\text{max}} \sim \eta \cdot k_{L_\text{max}}$. The absolute positions of these maxima and their position relative to each other do not depend on the Afv\'enic Mach number ($M_\text{A} = v_0 / v_\text{A}$) of the flow, if it is higher than $\sim 10$ (for $M_\text{A} \lesssim 5$, the growth rates are modified). It is shown by simulations~\cite{rpbsw} that ions passing through such a sub-resonant R-wave do not interact significantly with it; they instead gain a small wiggling motion around the equilibrium position along their flow. Even if the wavelength is set to resonant, interaction still remains very weak. For the wave amplitudes of $\sim B_0$, polarization of the wave is therefore a decisive property; it determines whether or not interaction with ions will emerge.

\begin{figure*}[ht!]
	\includegraphics[width=0.24\linewidth]{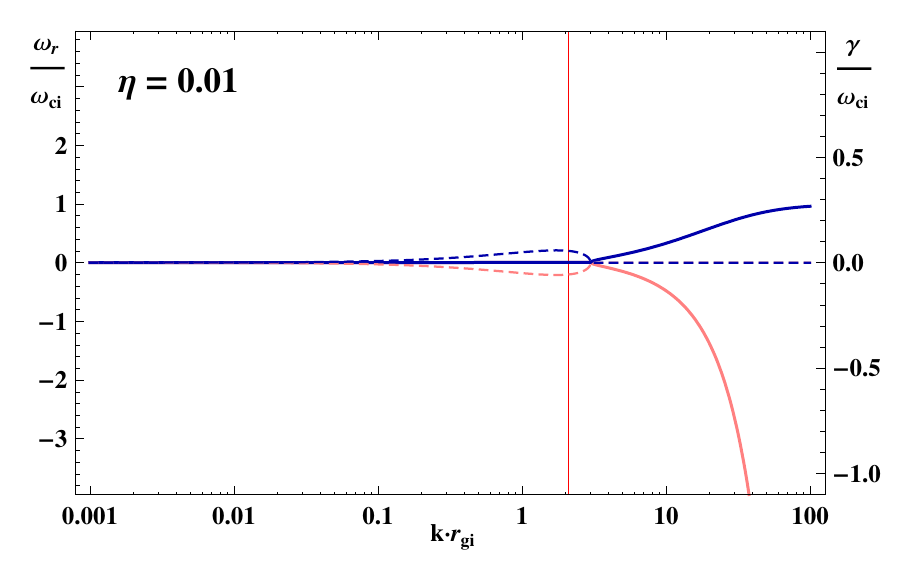}
	\includegraphics[width=0.24\linewidth]{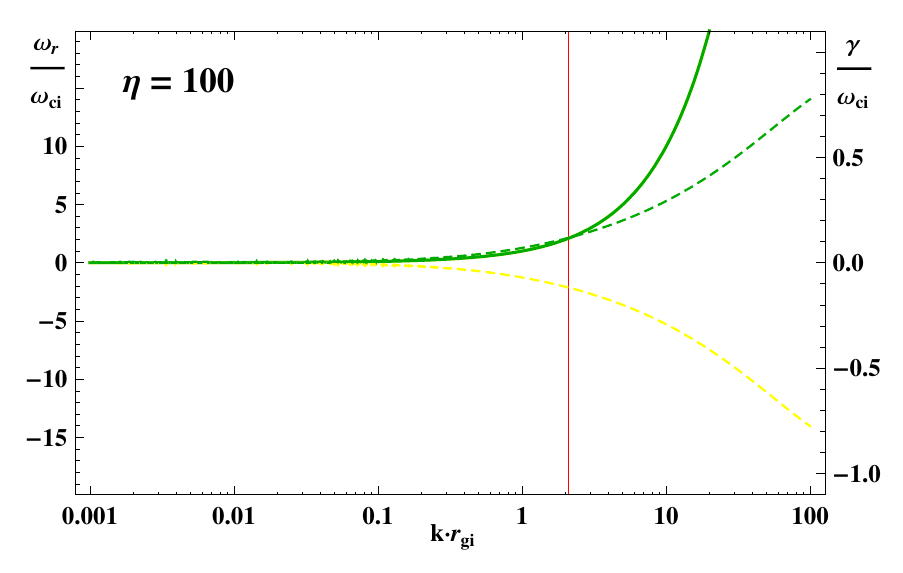}
	\includegraphics[width=0.24\linewidth]{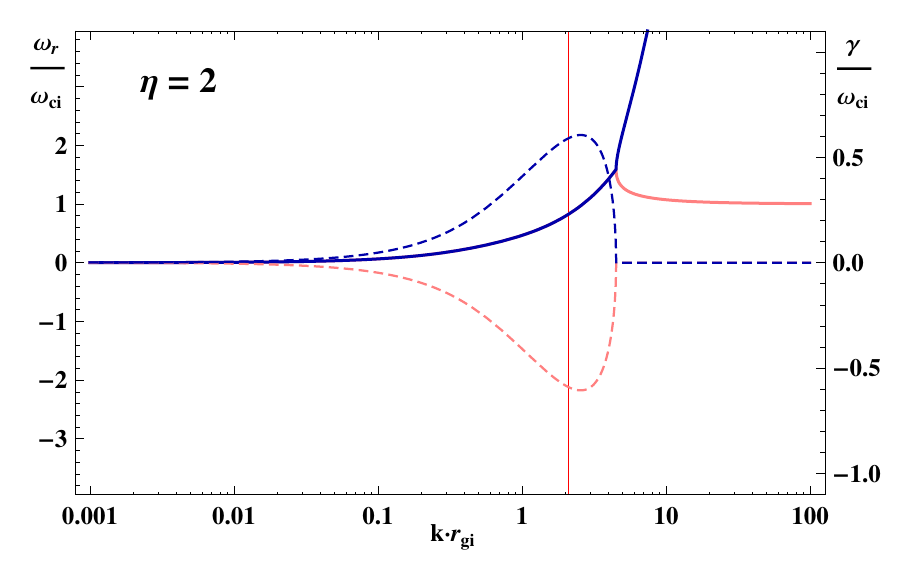}
	\includegraphics[width=0.24\linewidth]{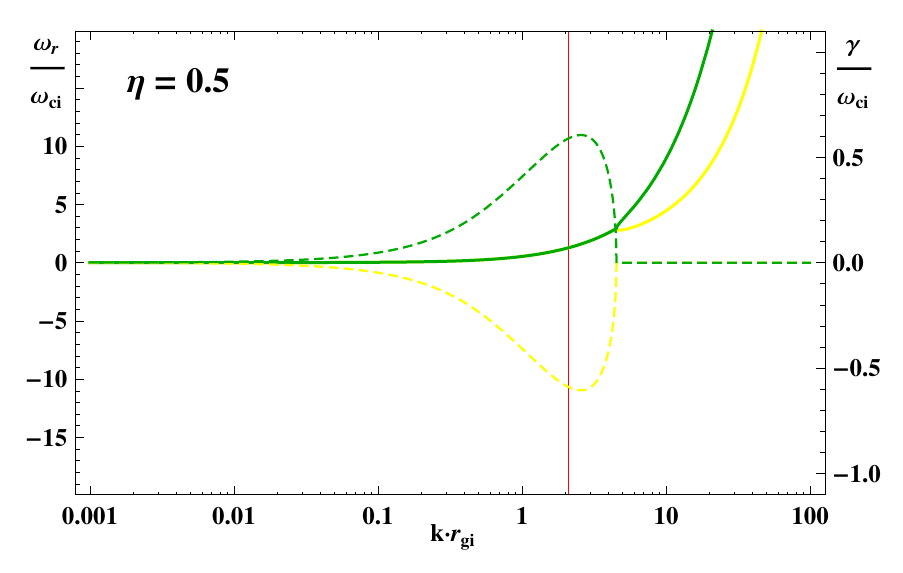}
	\includegraphics[width=0.24\linewidth]{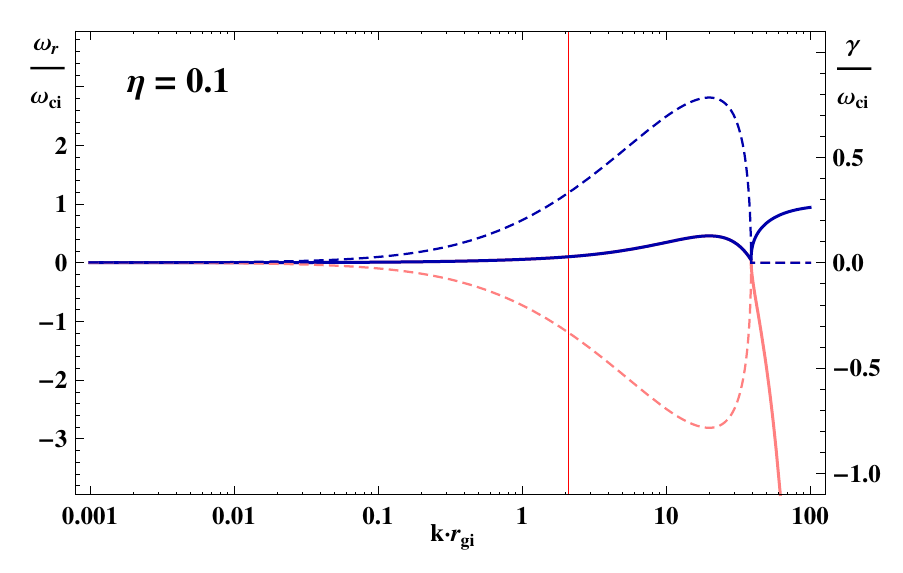}
	\includegraphics[width=0.24\linewidth]{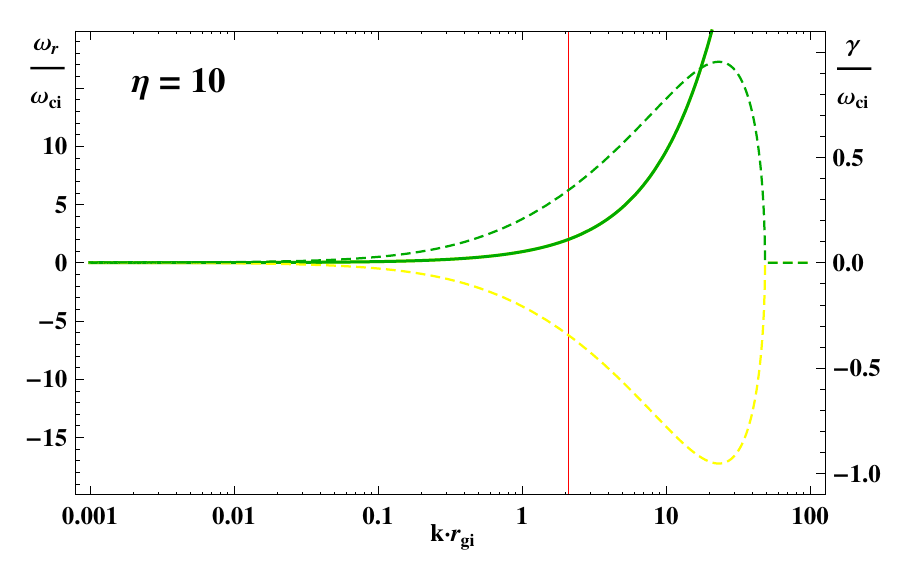}
	\includegraphics[width=0.24\linewidth]{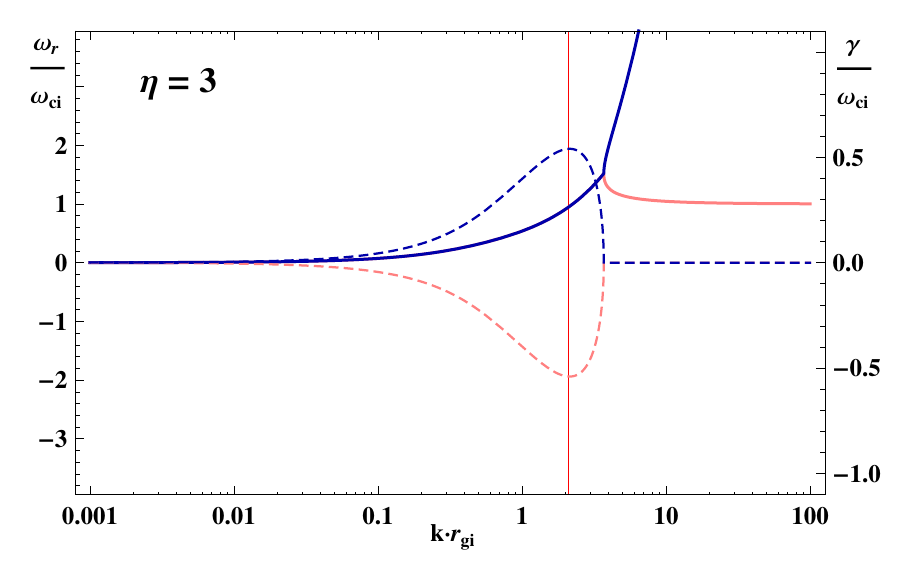}
	\includegraphics[width=0.24\linewidth]{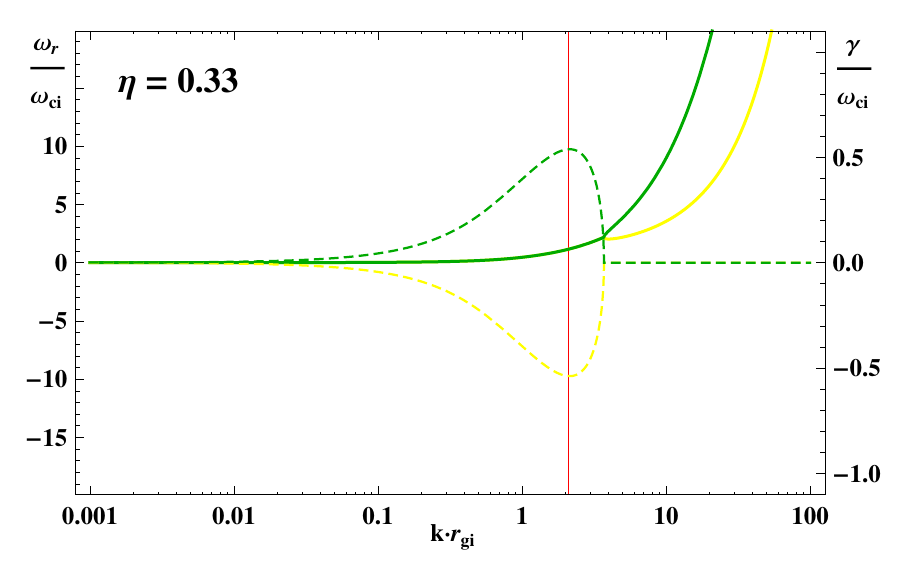}
	\includegraphics[width=0.24\linewidth]{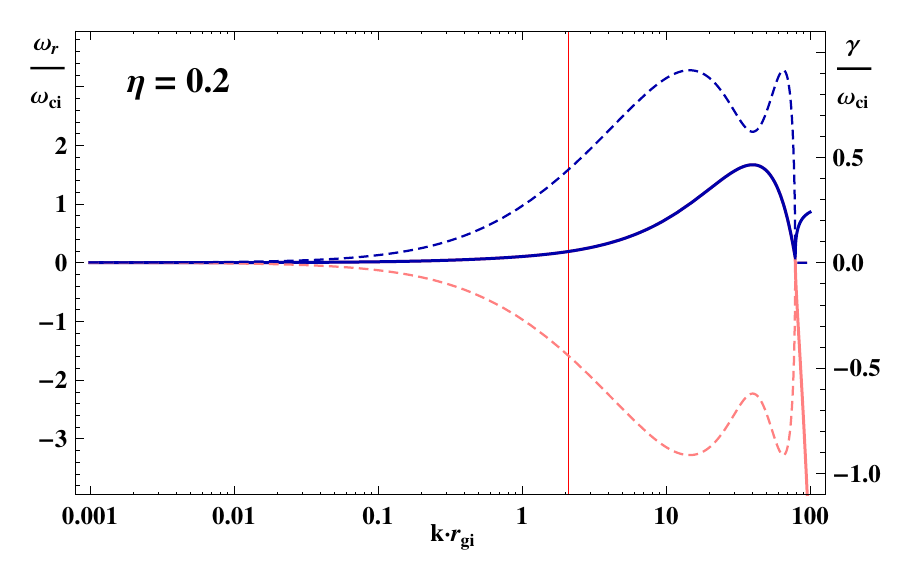}
	\includegraphics[width=0.24\linewidth]{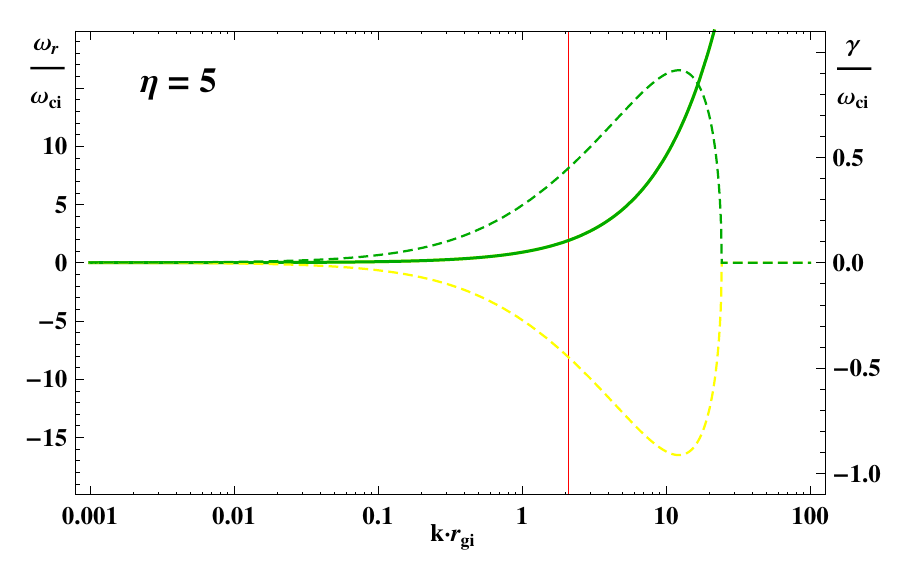}
	\includegraphics[width=0.24\linewidth]{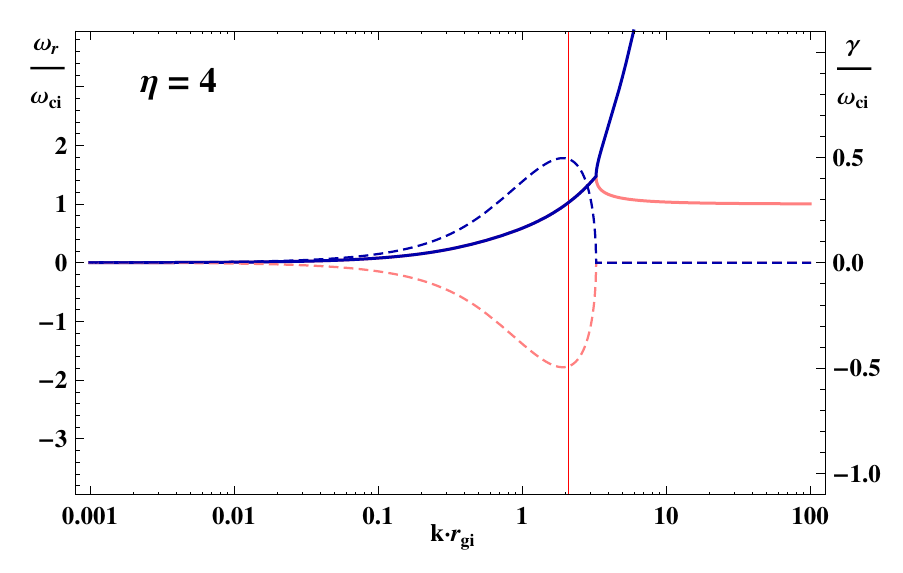}
	\includegraphics[width=0.24\linewidth]{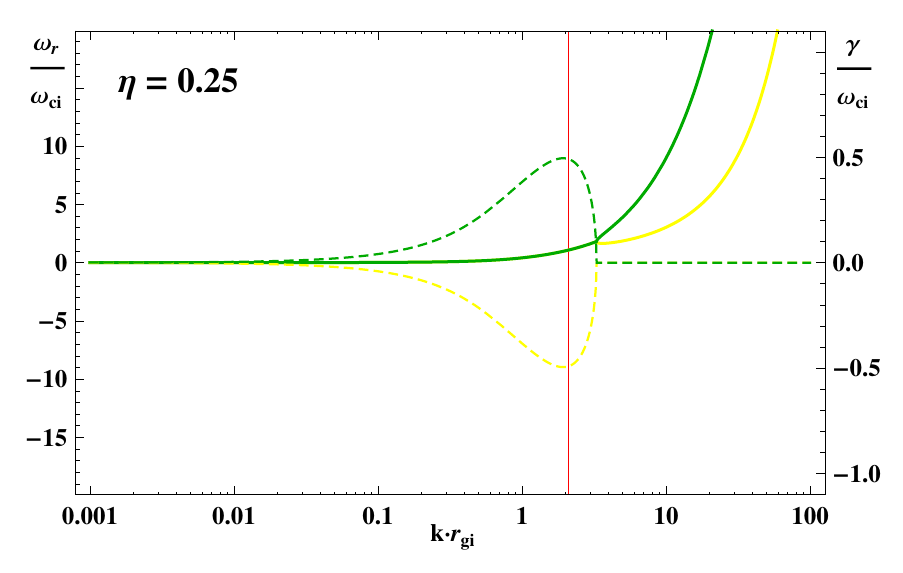}
	\includegraphics[width=0.24\linewidth]{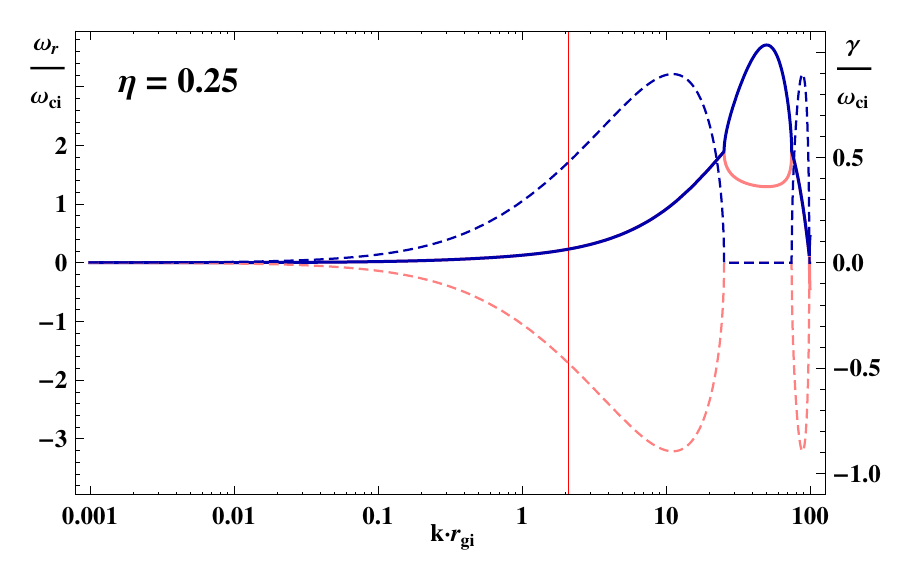}
	\includegraphics[width=0.24\linewidth]{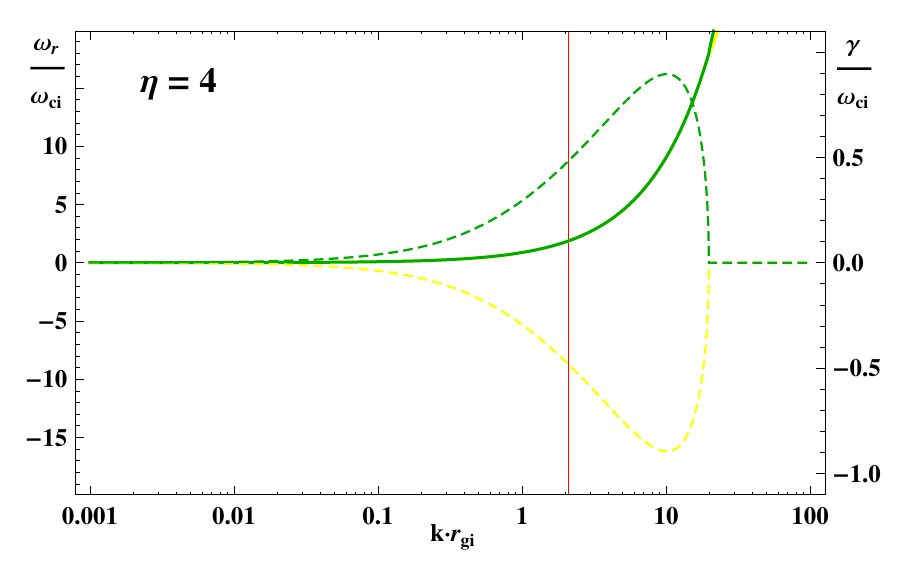}
	\includegraphics[width=0.24\linewidth]{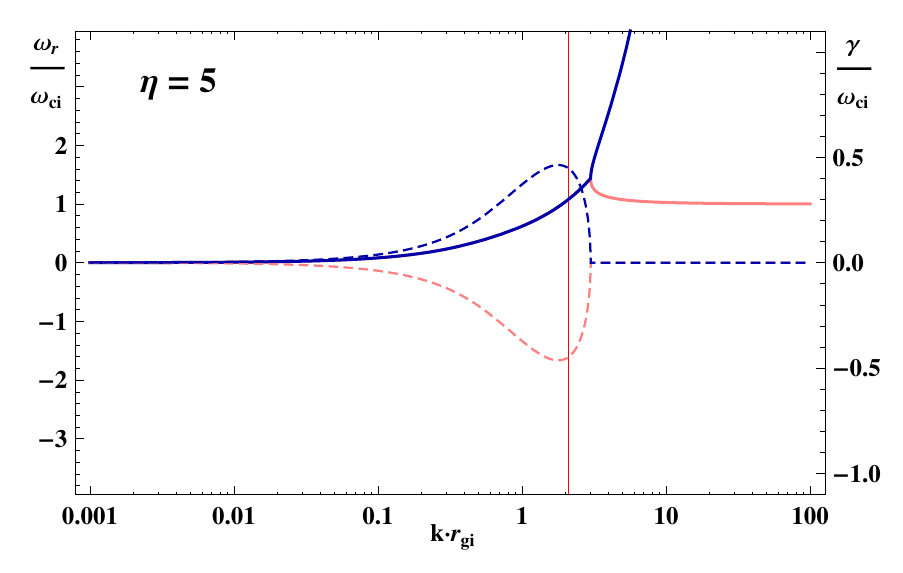}
	\includegraphics[width=0.24\linewidth]{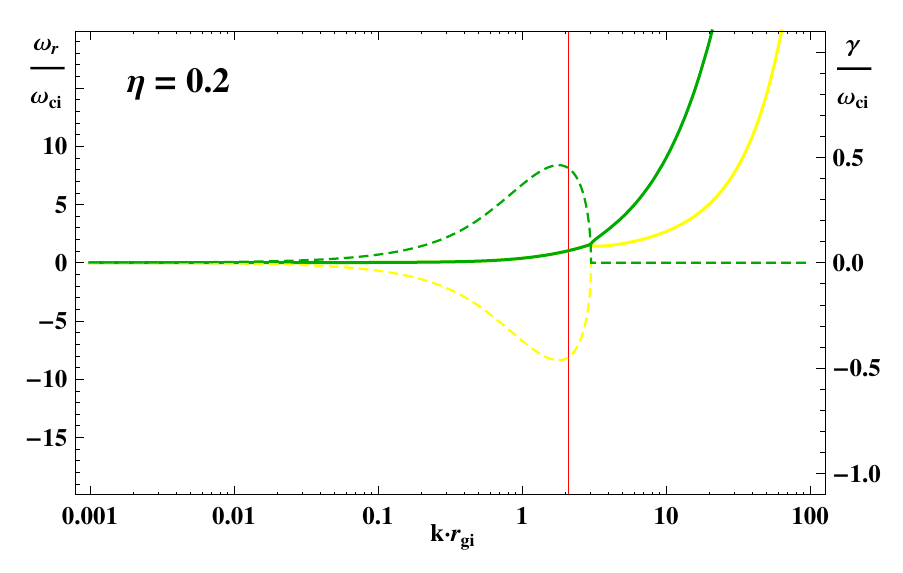}
	\includegraphics[width=0.24\linewidth]{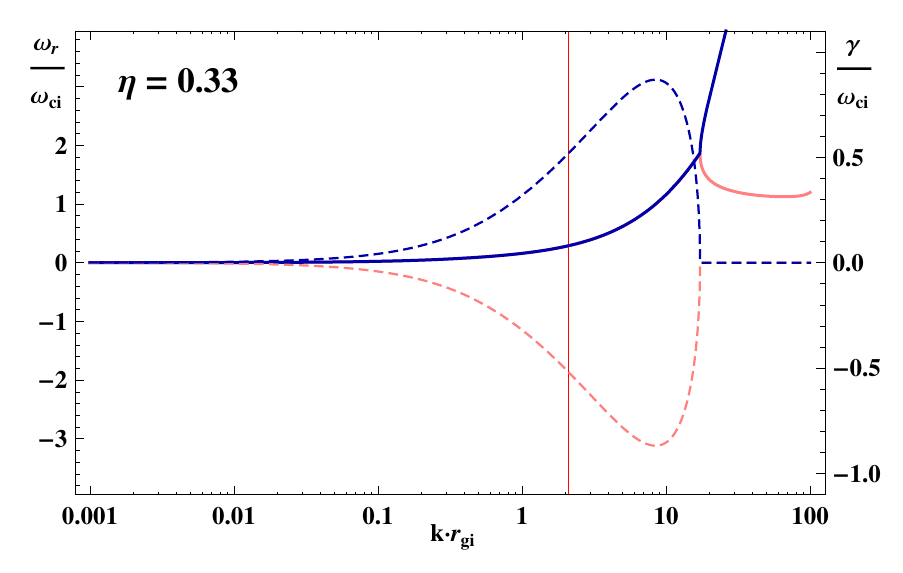}
	\includegraphics[width=0.24\linewidth]{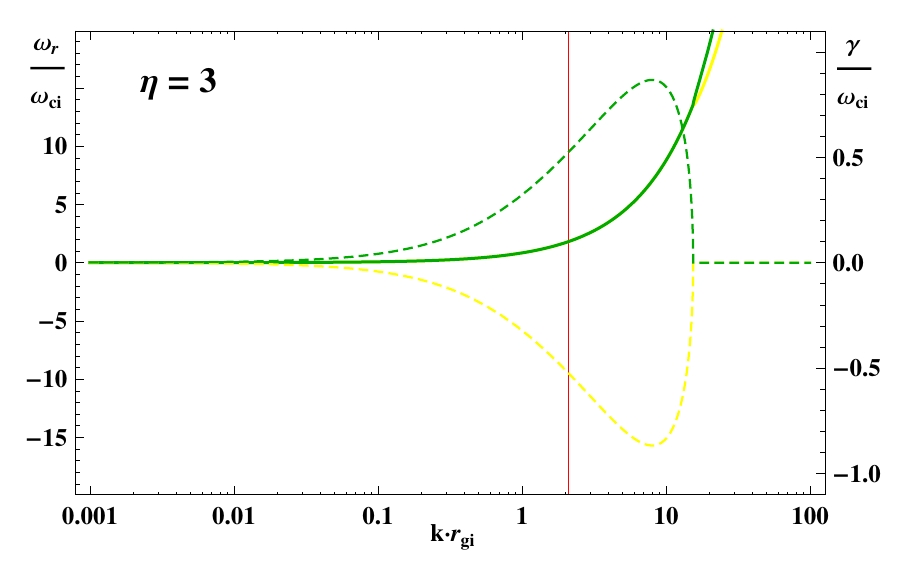}
	\includegraphics[width=0.24\linewidth]{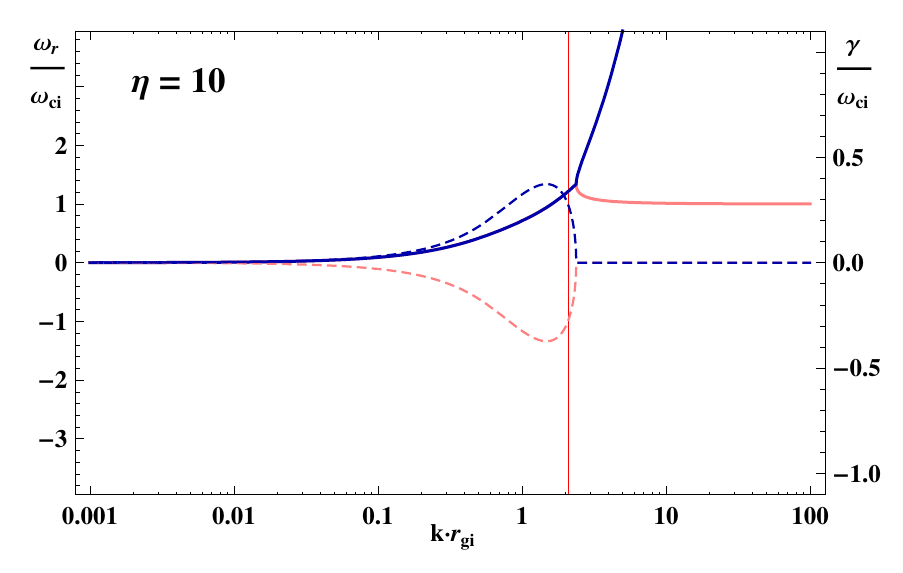}
	\includegraphics[width=0.24\linewidth]{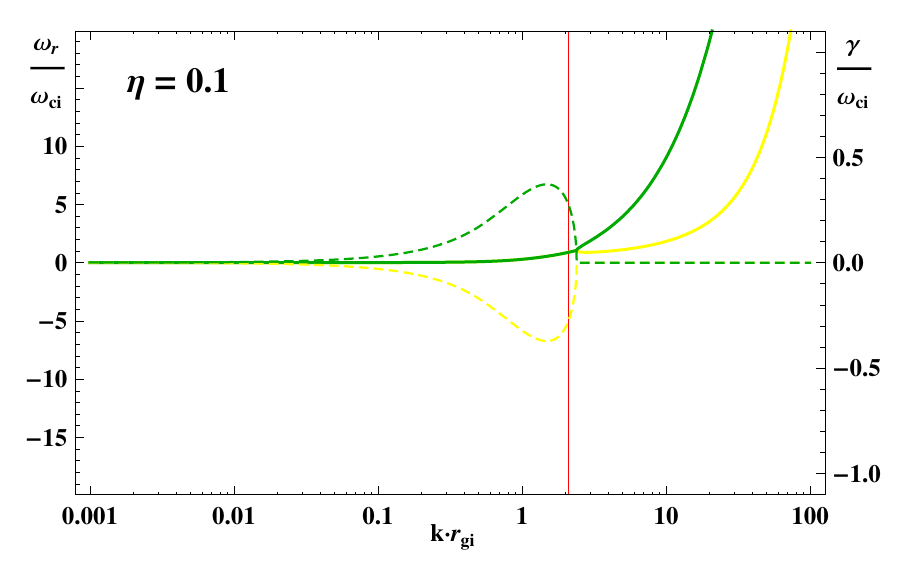}
	\includegraphics[width=0.24\linewidth]{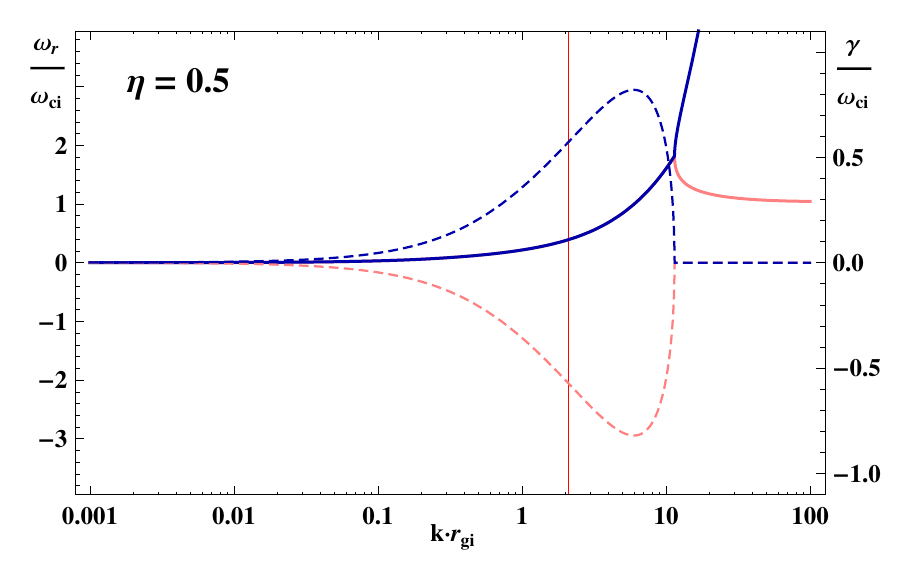}
	\includegraphics[width=0.24\linewidth]{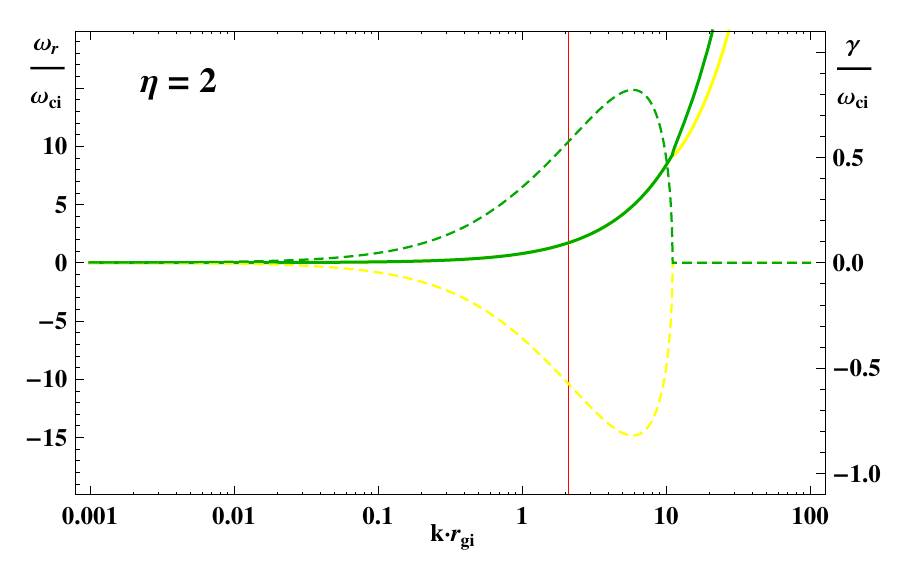}
	\includegraphics[width=0.24\linewidth]{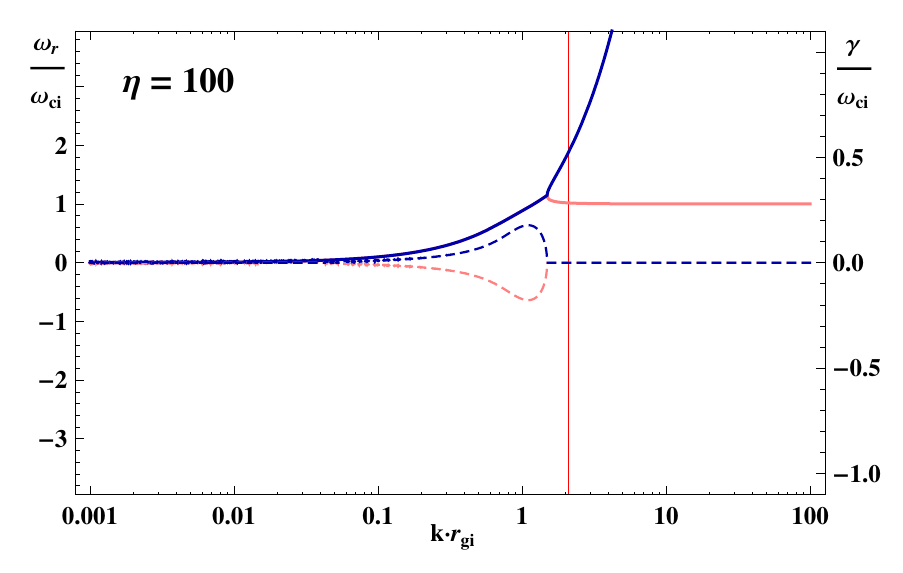}
	\includegraphics[width=0.24\linewidth]{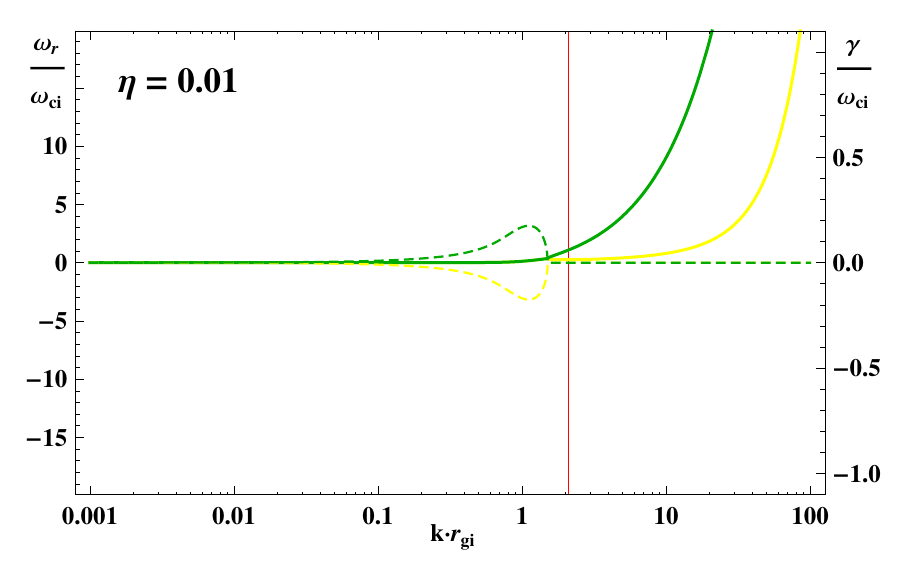}
	\includegraphics[width=0.24\linewidth]{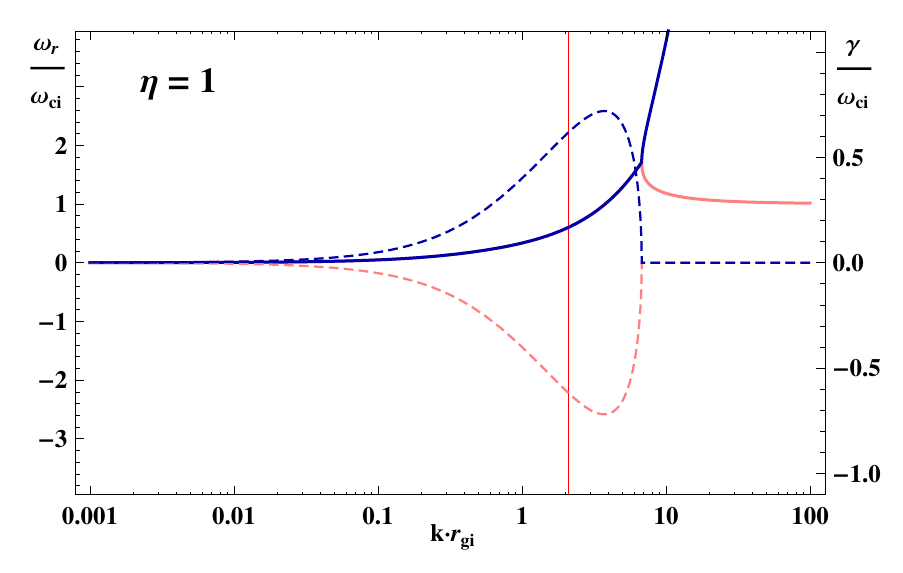}
	\includegraphics[width=0.24\linewidth]{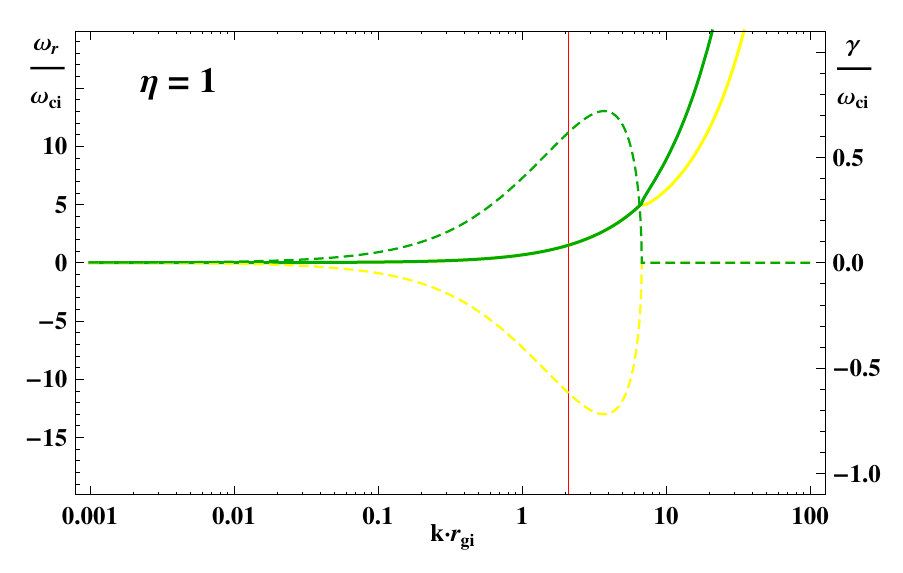}
	
	\caption{\label{fig:RLicomp} The L-wave (blue and red) and the R-wave (green and yellow) real frequency and growth rate dependences on wavenumber. Graphs of the L and R modes are sorted in pairs (L,~R), for which the product of their density ratios is $\eta_L \cdot \eta_R = 1$. The real part ($\omega_r$) of each of the roots is plotted as a continuous line and the imaginary part ($\gamma$) as a dashed line; both expressed in relative units of $\omega_{ci}$, while the wavenumber is given in units of $r_{gi}^{-1}$.}
\end{figure*}

In Fig.~\ref{fig:RLicomp} we see that for most of the values of $\eta$, the growing modes of L and R-waves at $\lambda_\text{res}$ have nearly the same growth rate. They have slightly different real frequencies and, therefore, different phase speeds. In the range of $\eta \sim [0.2 - 5]$, the growth rate of the resonant mode is $\gamma \approx 0.55~\omega_{ci}$, and the difference between the real frequencies of the two modes is in the range $\Delta \omega_{r} / \omega_{ci} \sim [0.5 - 1]$. When two modes are combined, the electric field vector of the superposed wave is found as

\begin{eqnarray}
	&&{\bf E}_L = E(\gamma t) e^{i({\bf k} \cdot {\bf x} - \omega_r t)} \frac{1}{\sqrt{2}} \begin{pmatrix}
		0 \\
		1  \\
		i \\
	\end{pmatrix}, \nonumber
	\\
	&&{\bf E}_R = E(\gamma t) e^{i({\bf k} \cdot {\bf x} - \omega_r t)} \frac{1}{\sqrt{2}} \begin{pmatrix}
		0 \\
		1  \\
		i \\
	\end{pmatrix} e^{-i \Delta \omega_r t}, \nonumber
	\\
	&&{\bf E}_{L+R} = E(\gamma t) e^{i({\bf k} \cdot {\bf x} - \omega_r t)} \frac{1}{\sqrt{2}}\begin{pmatrix}
		0 \\
		1+e^{-i \Delta \omega_r t}  \\
		i (1-e^{-i \Delta \omega_r t}) \\
	\end{pmatrix}. \nonumber
\end{eqnarray}

\noindent After multiplying the last expression by $e^{-i \Delta \omega_r t / 2} / e^{-i \Delta \omega_r t / 2}$, we apply the Euler's formula to get:

\begin{eqnarray}
	{\bf E}_{L+R} =~&&2 E(\gamma t)~e^{i [ {\bf k} \cdot {\bf x} - (\omega_r + \tfrac{\Delta \omega_r}{2}) t ]} \cdot
	\label{eq:linpol}
	\\
	&&\cdot~\frac{1}{\sqrt{2}}
	\begin{pmatrix}
		0 \\
		\cos \left(\frac{\Delta \omega_r t}{2} \right) \\
		-\sin \left(\frac{\Delta \omega_r t}{2} \right) \\
	\end{pmatrix}. \nonumber
\end{eqnarray}

In the linear regime, both waves grow at the same rate $\gamma$. Eq.~(\ref{eq:linpol}) implies that the superposed wave should grow as linearly polarized. Its plane of polarization should rotate (the Faraday rotation). We estimate the rotation period $\Delta T$ of the plane of polarization, by equating the phase of the rotation to $2 \pi$. From the previous equation we get:

\begin{equation}
	\frac{\Delta \omega_r}{2} \tau_\text{rot} = 2 \pi,~\tau_\text{rot} = 2 \cdot \frac{2 \pi}{\Delta \omega_r}. \nonumber
\end{equation}

For the resonant wavenumbers ($\Delta \omega_r \approx \omega_{ci}$), the rotation period is $\tau_\text{rot} \approx 2 \cdot T_{ci}$, where $T_{ci}$ is the period of time required for an ion to make one orbit around the magnetic field line. Due to a fast growth rate, the wave remains in the linear regime only for a few cycles, until the amplitude grows enough ($\sim B_0$) to trigger the resonant scattering of ions~\cite{rpbsw}. We think that in such a non-linear regime, the magnetic field line bends to a helicoidal shape, and the vector of the electric field starts to rotate. Since the R-wave only weakly interacts with ions~\cite{rpbsw}, this will eventually become the point of separation of the L and R modes, where polarization of the wave turns from a rotating linear to a left circular.

We emphasize here that in case of the cold and weak ($\eta \lesssim 0.1$) ion beam, the maximum of the growth rate for the R and L modes shifts toward the lower wavenumbers. For the R-mode, the position of the maximum is then near the point where $k \cdot r_{gi} \approx 1$. The obtained real frequency and the growth rate have the same values as those studied in~\cite{microinst,emibinst} for the ion/ion right-hand resonant instability. The cyclotron resonant condition $\omega_r \simeq {\bf v}_0 \cdot {\bf k} \pm \omega_{ci}$ then applies too. For a fixed Alfv\'enic Mach number, lowering the density of the flow makes the L-wave instability turn off much earlier than the R-wave instability. The L-mode instability of such a tenuous beam has the same value and shape of $\omega_r$ and $\gamma$ dependences, as in the case of the ion/ion non-resonant instability~\cite{microinst,emibinst}. However, here, this mode propagates in the direction of the flow. This is probably the reason this instability is left circularly polarized, instead of being right circularly polarized and propagating opposite to the flow, as studied in~\cite{microinst,emibinst}.

In Fig.~\ref{fig:RLicomp}, the R and L modes are shown together in pairs for the different density ratios of the two plasmas. Besides the double peak that emerges at ratios $\eta \sim 0.2 - 0.3$ of the L-mode, we observe the global symmetry between the growth rates of these modes. However, their real-frequency solutions differ in amplitude.

\subsection{\label{sec:normal}Wavevector normal to the magnetic field}

Without loss of generality, we take the wavevector to be in the direction of $z$-axis. We find (see Sec.~\ref{sec:appnormal} in the Appendix) that the plasma flowing in the direction parallel to the magnetic field lines has no influence on extraordinary EM (X-mode) waves. However, it can make ordinary EM (O-mode) waves subjected to the Weibel instability. On the other hand, the plasma flowing in the direction perpendicular to the magnetic field lines and parallel to ${\bf k} = k {\bf \hat{z}}$ has no influence on the O-mode, but it can make the X-mode become unstable.

If the flow is perpendicular to ${\bf B}_0$ and parallel to the wavevector (${\bf v}_0 = v_0 {\bf \hat{z}}$), the polarization matrix given by Eq.~(\ref{eq:gdm}) takes the form:

\begin{eqnarray}
	&&\mathcal{D} =
	\begin{pmatrix}
		P - n^2 & 0 & 0 \\
		0 & S - n^2 & -i D \\
		0 & i D & S \\
	\end{pmatrix} + \label{eq:mdmnp}
	\\
	&&+ \sum_{\alpha = i,e} \frac{\omega_{p\alpha f}^2}{\omega^2} \cdot
	\begin{pmatrix}
			~~0~ & 0 & 0 \\
			~~0~ & 0 & - i \dfrac{\omega_{c\alpha} k v_0}{\xi^2 \omega^2 - \omega_{c\alpha}^2} \\
			~~0~ & i \dfrac{\omega_{c\alpha} k v_0}{\xi^2 \omega^2 - \omega_{c\alpha}^2} & -\dfrac{ \omega k v_0 }{\xi^2 \omega^2 - \omega_{c\alpha}^2}(\xi + 1) \\
	\end{pmatrix}. \nonumber
\end{eqnarray}

The X-mode dispersion relation is given by the root:

\begin{equation}
	(S - n^2) S_M - D_M^2 = 0,~n^2 = \frac{R L - S M_S - 2 D M_D - M_D^2}{S - M_S}, \nonumber
\end{equation}

\noindent where $S_M$ and $D_M$ are derived in Sec.~\ref{sec:appnormal} in the Appendix. For frequencies around the ion-cyclotron or lower, the dispersion relation remains complicated and we solve it numerically.

\begin{figure*}[ht!]
	\includegraphics[width=0.25\linewidth]{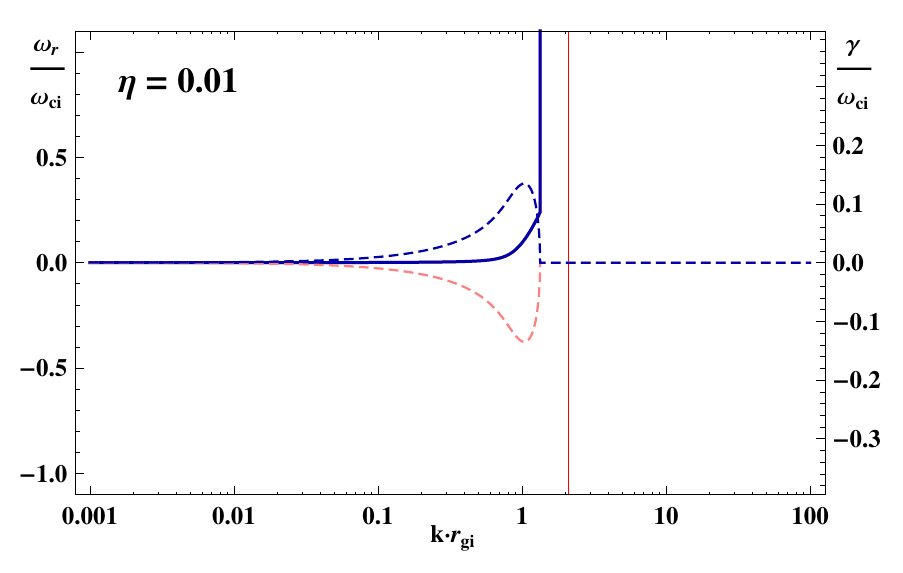}
	\includegraphics[width=0.25\linewidth]{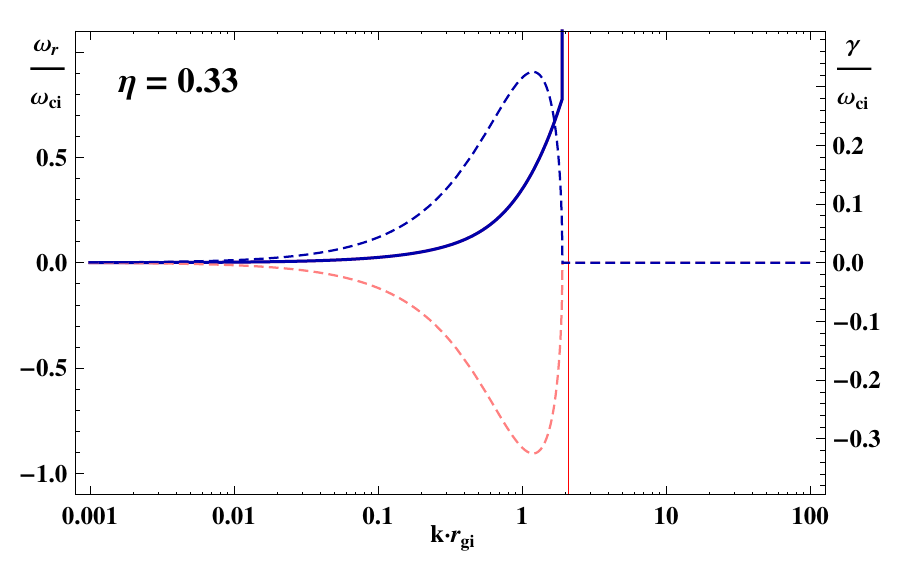}
	\includegraphics[width=0.25\linewidth]{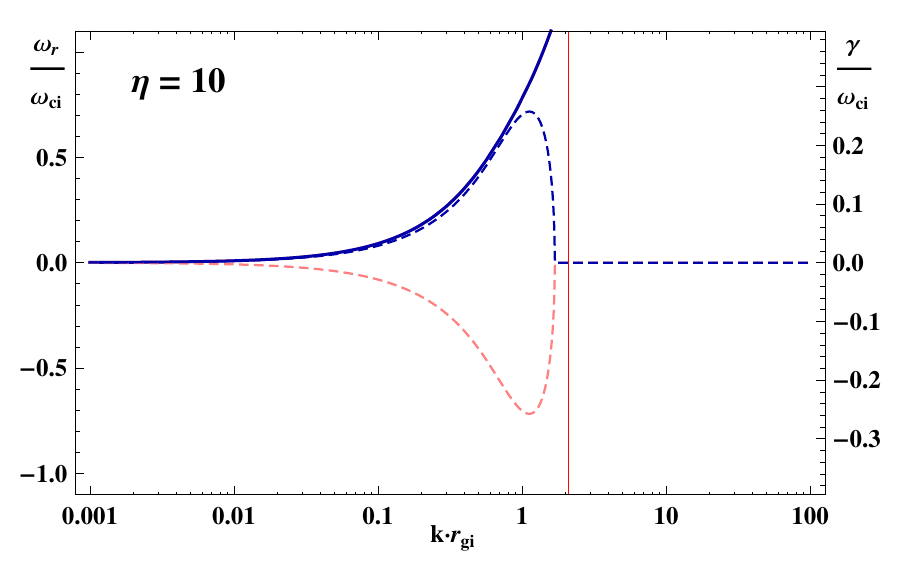}
	\includegraphics[width=0.25\linewidth]{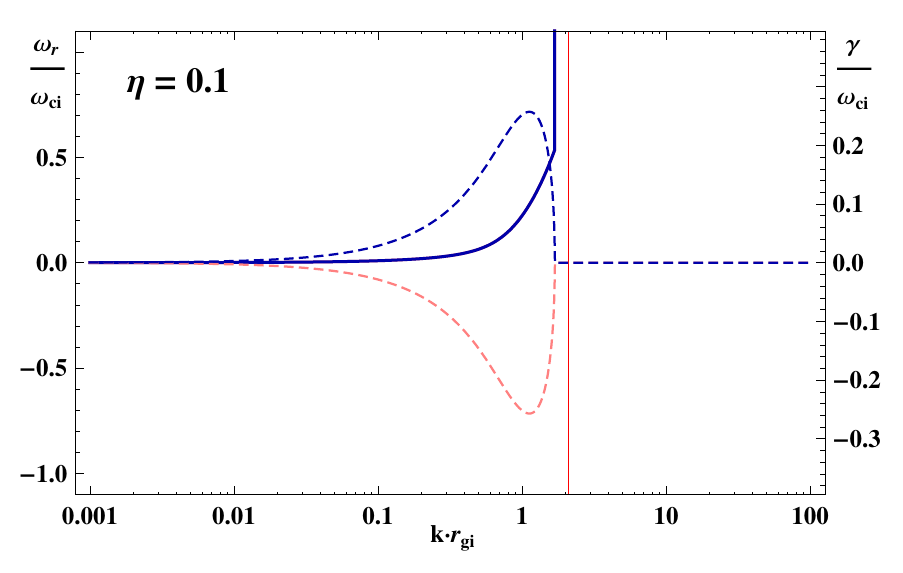}
	\includegraphics[width=0.25\linewidth]{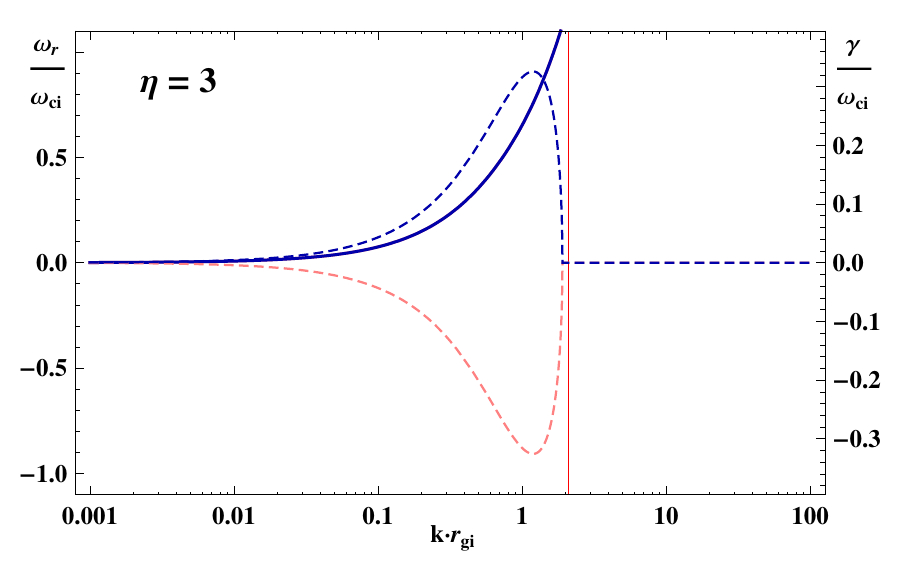}
	\includegraphics[width=0.25\linewidth]{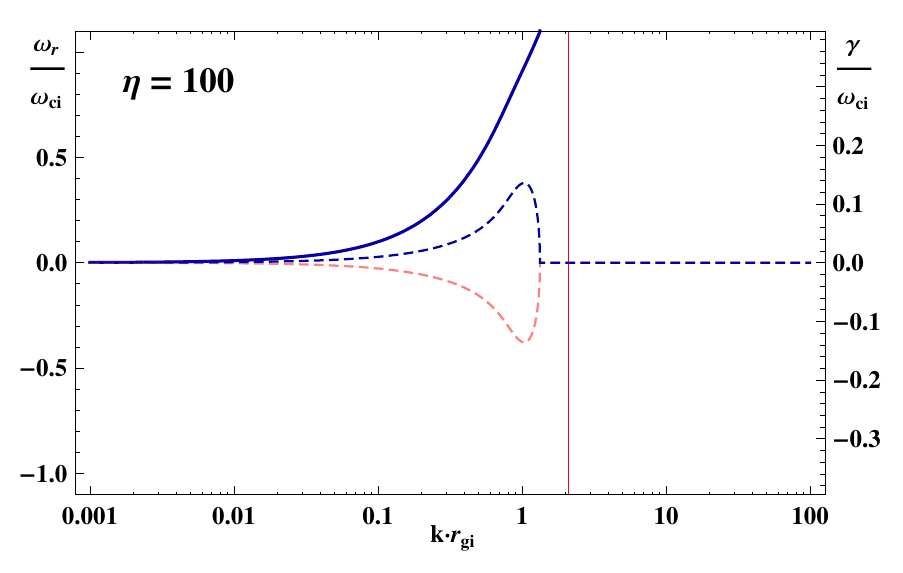}
	
	\caption{\label{fig:Xicomp} Similar to Fig.~\ref{fig:RLicomp}, only for the X-wave instability. The curves of the growing and decreasing waves are given in dark blue and red color, respectively. The position of the maximum is near the resonant wavenumber, at $k_{X_\text{max}} = 2 \pi / \lambda_\text{res} \approx 1/r_{gi}$.}
\end{figure*}

Similarly to parallel modes of propagation, the modified X-mode also becomes unstable if the flow is super-Alfv\'enic. The difference here is that the wavenumber of the maximum growth of the X-mode instability is somewhat lower ($\sim r_{gi}^{-1}$) than that of the R and L modes. The maximum growth rate is also at least two times smaller for this mode. The instability of the perpendicular flow is presented for different $\eta$ in Fig.~\ref{fig:Xicomp}. Contrary to the case of the parallel modes, the position of the maximum for the X-mode does not depend on the velocity of the flow. Nevertheless, its growth rate does.

Finally, if the plasma flow is inclined to the background magnetic field, the resulting instability is a composite one. Its component parallel to $B_0$ is composed of the R or L mode, and the component which is normal to $B_0$ consists of the X-mode.

\section{\label{sec:pic}Particle-In-Cell simulations of shock waves}

In this section we use the kinetic simulations to resolve the whole process of a quasi-parallel collisionless shock formation in a fully non-linear case. We emphasize the potential role of resonant micro-instabilities in a shock triggering. Once the shock is developed, we show how the same type of instability is excited in the upstream by a return plasma beam. We show how this upstream micro-instability, once it becomes non-linear, governs the process of shock reformation.

\subsection{\label{sec:insnonlin}Instability in the non-linear regime}

We first consider the wave-plasma interaction that eventually leads to the triggering of a shock wave. To understand what happens in the initial phases of the non-linear regime, we combine the linear theory and the results of non-linear test particle simulations~\cite{rpbsw}. In this estimation, we consider the interaction in a laboratory frame, which is the frame of the ISM. We assume that the excited wave in the non-linear regime does not steepen significantly. For the monochromatic transverse wave, it is shown in~\cite{Sagdeev} that the energy will not be significantly transferred from fundamental to higher harmonics. This means that the waves that deviate from the linear dispersion law are not much distorted due to non-linearity.

Static plasma (ISM) is reflected off the moving ejecta, so initially the two interpenetrating plasmas have the same densities ($\eta = 1$). If the velocity of the flowing plasma is $v_0 \gtrsim v_\text{A}$, the fastest growing mode is the resonant one (see Fig.~\ref{fig:RLicomp}), which has $\lambda = \frac{1}{2} \pi r_{gi}$ ($k = 4 / r_{gi}$). For a very short period of time, the instability should grow as a linearly polarized wave that propagates with a phase velocity $v_{\text{ph}} \approx \frac{1}{2} v_0$ in the direction of the flow. In the instability frame (the center-of-mass frame), the flowing plasma is moving with relative velocity $\Delta^f v \approx \frac{1}{2} v_0$. For this relative flow, the wavelength of the instability is resonant:

\begin{equation}
	\lambda \approx \frac{1}{2} \pi \frac{m_i v_0}{q_i B_0} = \pi \frac{m_i \Delta^f v}{q_i B_0} = \pi \Delta^f r_{gi}. \nonumber
\end{equation}

Therefore, we expect that ions are strongly scattered~\cite{rpbsw} once this wave reaches the amplitude $\sim B_0$. Because shocks are strongly non-linear, with this estimate we can only discuss the interaction during the period until the particles become scattered by the instability.

\subsection{\label{sec:simset}Simulation setup}

In a supernova explosion, the stellar material is blown away from the center of the star, and initially, it freely propagates through ISM. Due to a much larger density and amplified magnetic field, the expanding ejecta behaves like a physical barrier, reflecting the ISM particles it encounters. This process is frequently presented in 2D or 3D PIC simulations~\cite{tristan-mp,unmagnetized,rvsnr}, where the initiated plasma flow is simply reflected off the wall that replaces the ejecta. In the region where two counter-streaming beams overlap, instabilities develop and make the plasma become turbulent, forming a shock wave. The simulation frame thus corresponds to the downstream (or ejecta) frame.

We numerically solve kinetic plasma equations by using the PIC code TRISTAN-MP \cite{tristan-mp}. We initiate the shock by reflecting a cold stream of particles (ions + electrons coming from the right side of a simulation box) from a conducting wall that is on the left. In one case, we consider the background magnetic field lines that are exactly parallel to the flow. In the other case, the lines are inclined relative to the direction of the flow with the inclination angle $\theta = 15^{\circ}$. Simulation setup here is similar to the case presented in \cite{sda}. We use a fixed size simulation box of a rectangular shape in the $x-y$ plane, with periodic boundary conditions in the $y$ direction. The computational domain is 256 cells wide (along $y$) and 50000 cells long (along $x$), with the skin depth of 10 cells and each cell initially containing 4 particles (two electrons and two ions). In physical units, the size of the domain is $\sim 25 \times 5000~c / \omega_{pe}$. The end time of the simulation is $T = 4500~\omega_{pe}^{-1}$. The bulk Lorentz factor of the electron-ion flow is given by $\gamma_0$, and the beam thermal spread is $\Delta\gamma_0 = 10^{-8}$. Magnetization is defined as the ratio of magnetic to kinetic energy density $\sigma = B_0^2 / 4 \pi \gamma_0 n_i m_i c^2 = \omega_{ci}^2 / \omega_{pi}^2$. We choose the constant values for electron magnetization $\sigma_e = \omega_{ce}^2 / \omega_{pe}^2 = 0.1$ and the mass ratio $m_i / m_e = 16$, which implies $\sigma = 0.00625$ and $v_\text{A} \approx 0.08~c$. All the simulation runs are restricted to low $M_\text{A} < 10$.

\subsection{\label{sec:picd}Resonant instability and shock triggering}

\begin{figure*}[!ht]
	\centering
	\includegraphics[width=0.9\textwidth,height=0.07\textheight]{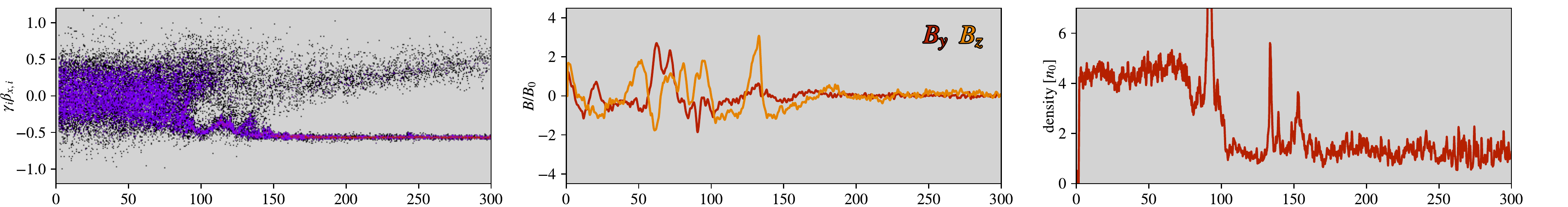}
	\includegraphics[width=0.9\textwidth,height=0.07\textheight]{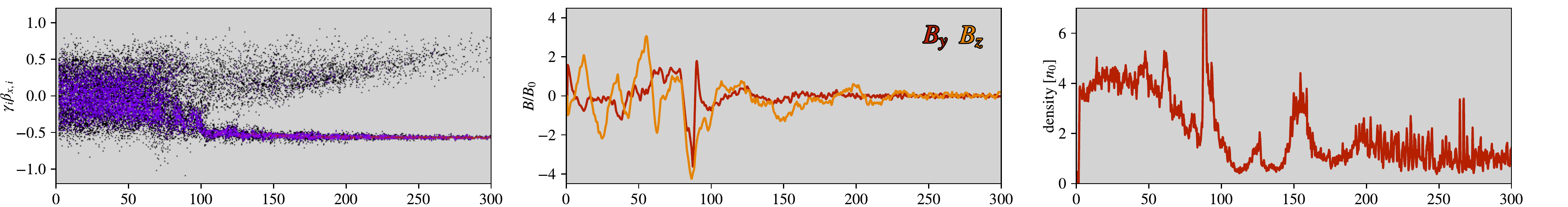}
	\includegraphics[width=0.9\textwidth,height=0.07\textheight]{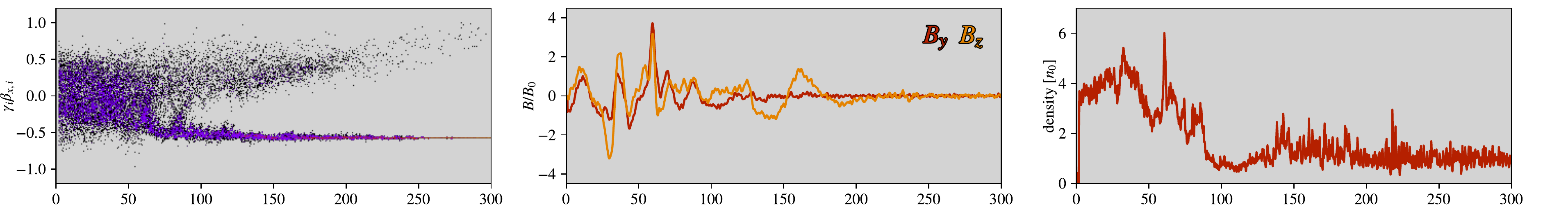}
	\includegraphics[width=0.9\textwidth,height=0.07\textheight]{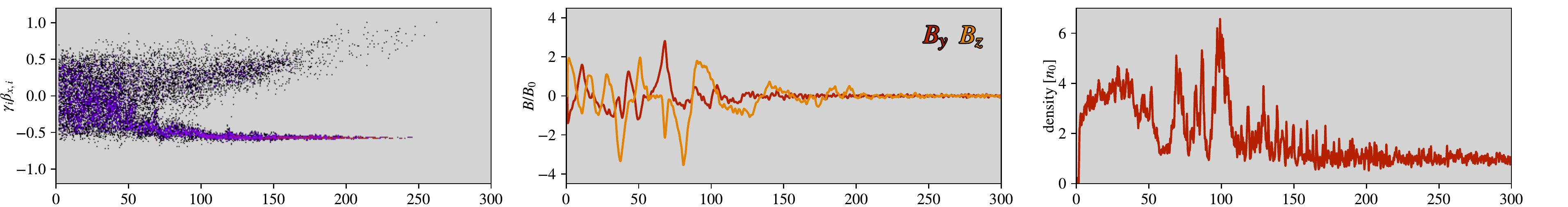}
	\includegraphics[width=0.9\textwidth,height=0.07\textheight]{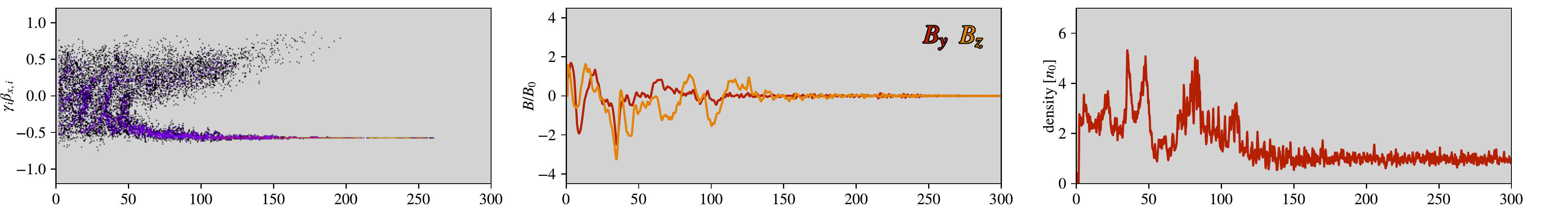}
	\includegraphics[width=0.9\textwidth,height=0.07\textheight]{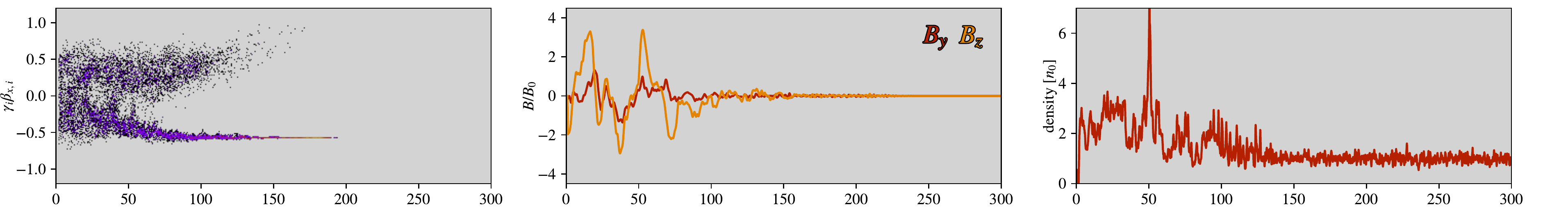}
	\includegraphics[width=0.9\textwidth,height=0.07\textheight]{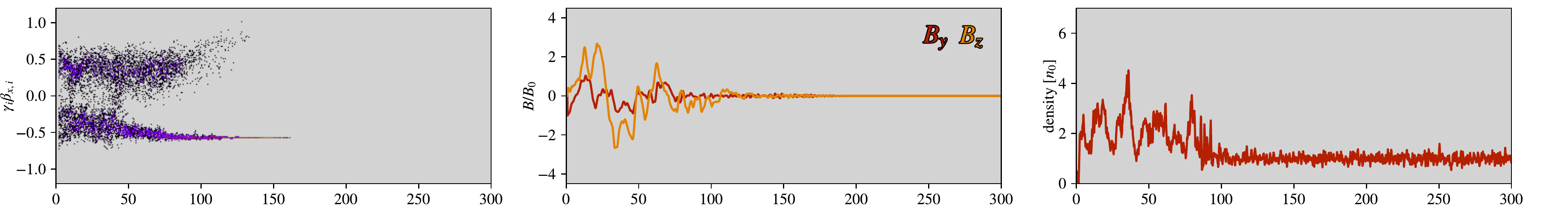}
	\includegraphics[width=0.9\textwidth,height=0.07\textheight]{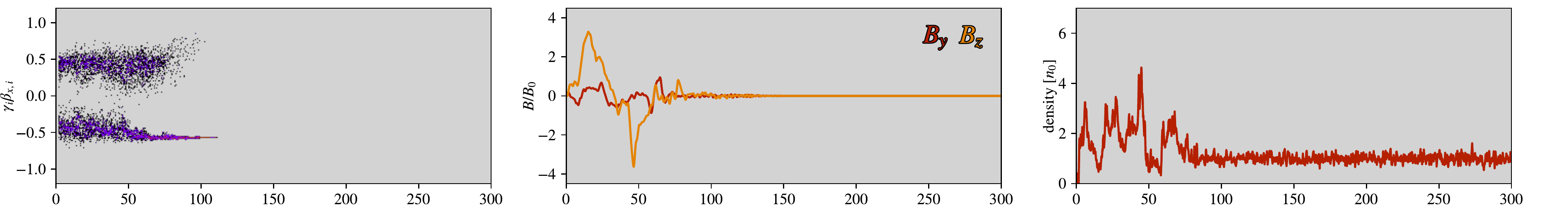}
	\includegraphics[width=0.9\textwidth,height=0.07\textheight]{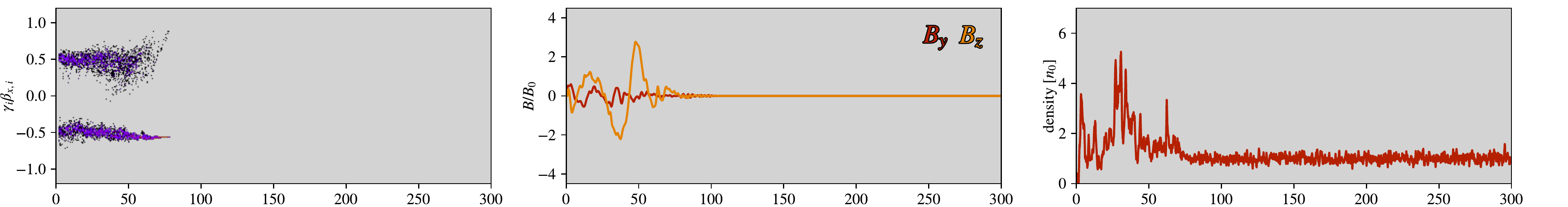}
	\includegraphics[width=0.9\textwidth,height=0.07\textheight]{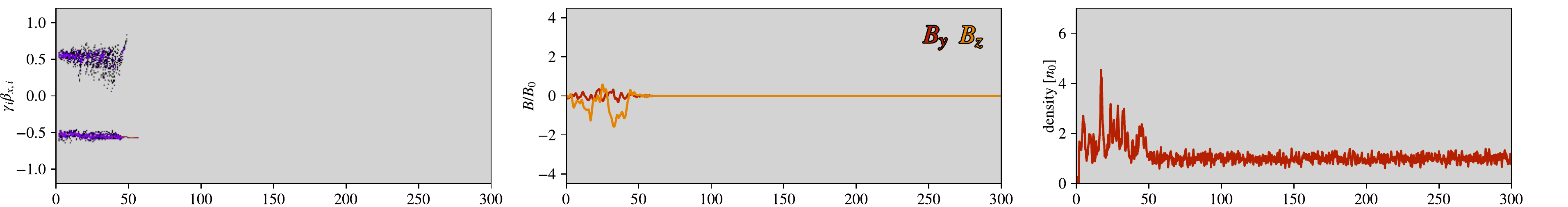}
	\includegraphics[width=0.9\textwidth,height=0.07\textheight]{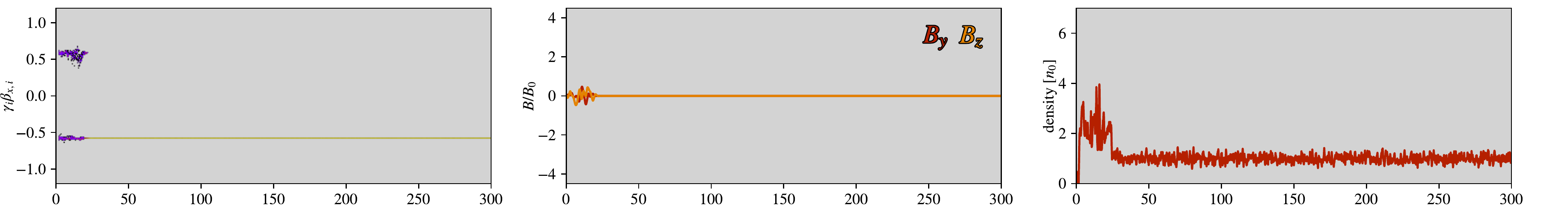}
		
	\caption{\label{fig:pic0} The $p_x$-$x$ phase space, transverse fields $B_y$ and $B_z$, and density profile captured at the very first phases of shock formation $t = 45 - 900 \ \omega_{pe}^{-1}$. On the horizontal axis, the \emph{x}-coordinates are given in units of $c / \omega_{pe}$. }
\end{figure*}

Here, we present the resonant mechanism of a shock triggering, which we observe in PIC simulations. Initially, a beam of particles is reflected off the conducting wall, and the two beams start to interpenetrate each other. Right at the first stage of the collision, we observe that the wave which is resonant with the velocity of the inflowing plasma is excited inside the overlapping region (see Fig.~\ref{fig:pic0}). Once the wave grows to the amplitude $\sim B_0$, we find that ions become strongly scattered by this wave. In the phase spectrum, we observe that the two counter-streaming beams start to scatter and merge periodically at localized regions. We see that the periodicity of this merging is related to the wavelength of the growing instability. Such behavior in the phase spectrum is expected, if the interaction is of the resonant type (see Fig.~2 in Ref.~\cite{rpbsw}). The wave grows as linearly polarized, and once it couples non-linearly with the plasma, its polarization switches to a left circular. We also find that its wavelength depends on the ion mass and the velocity of the flow as $\lambda \sim m_i v_0 / q_i B_0$, which reveals the Alfv\'enic nature of the wave.

By previous properties, we find that this wave strongly matches the strong-beam resonant instability that we analyzed in Sec.~\ref{sec:resalfinst}. In Fig.~\ref{fig:pic0}, we see  that the shock triggering occurs in the same way as we discussed in Sec.~\ref{sec:insnonlin}. The resonant wave therefore disrupts the flow by capturing the ions. We found that in this stage, the wave remains nearly monochromatic, and its shape is only slightly distorted. This is in part expected for the transverse monochromatic wave that enters the non-linear stage~\cite{Sagdeev}.

Immediately afterwards, ions that are not captured (the part of the beam slightly in front of the wave) are thus cut off by the resonant mode. This sparse bunch of ions continues to propagate and excites the wave of an opposite polarization ahead of the forming shock. This wave grows and advects towards the shock. In this stage, we observe a sharp change in the phase angle of the magnetic field, which clearly separates the two waves. The reformation is triggered once the upstream wave grows enough so that its amplitude can be compressed to at least $\sim 3-4~B_0$ at the shock. The shock reformation is further governed by the upstream wave, which completely overtakes the processes of particle thermalization and shock transmission. 

\begin{figure*}[!ht]
	\centering
	\includegraphics[width=0.49\linewidth]{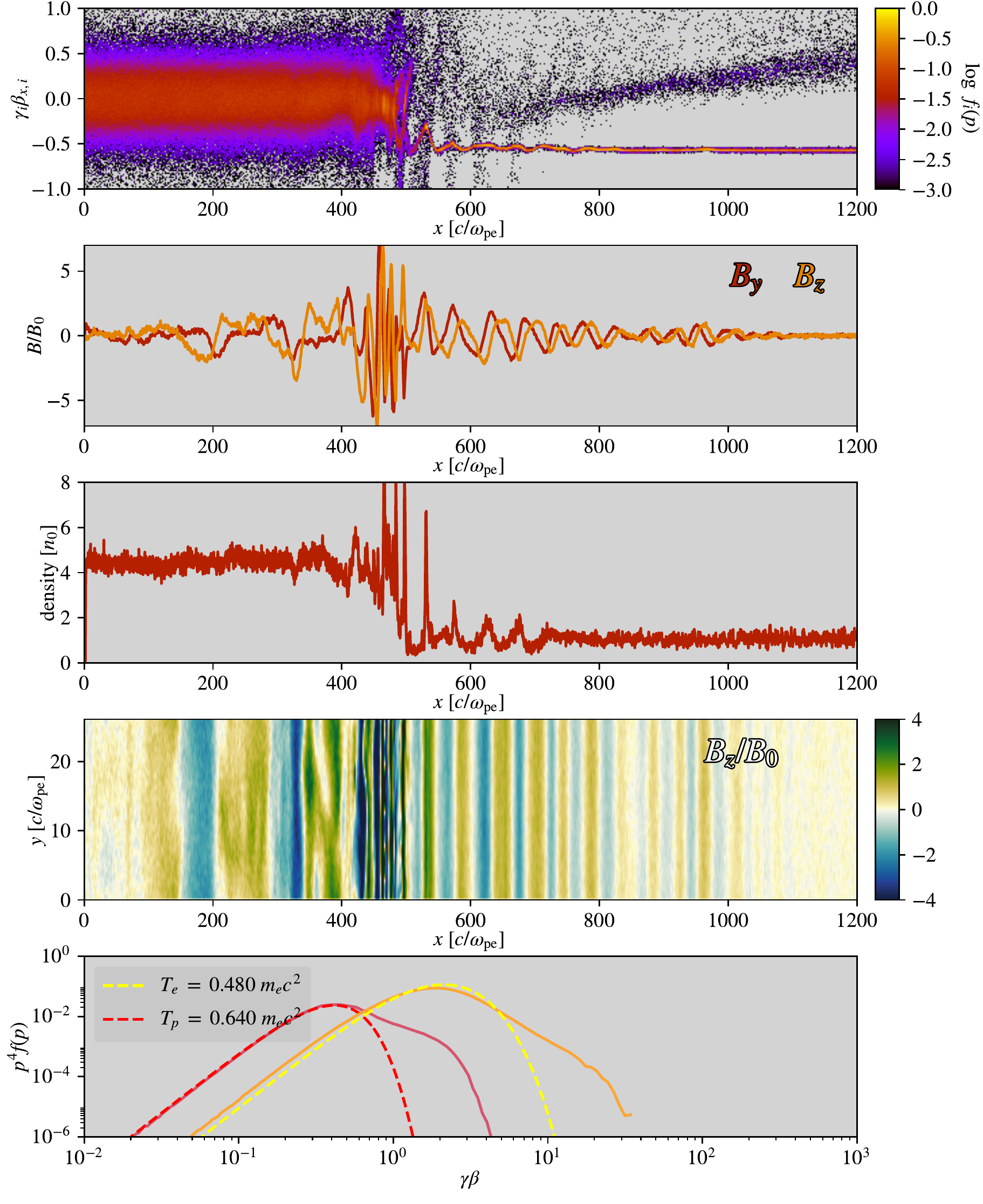}
	\includegraphics[width=0.49\linewidth]{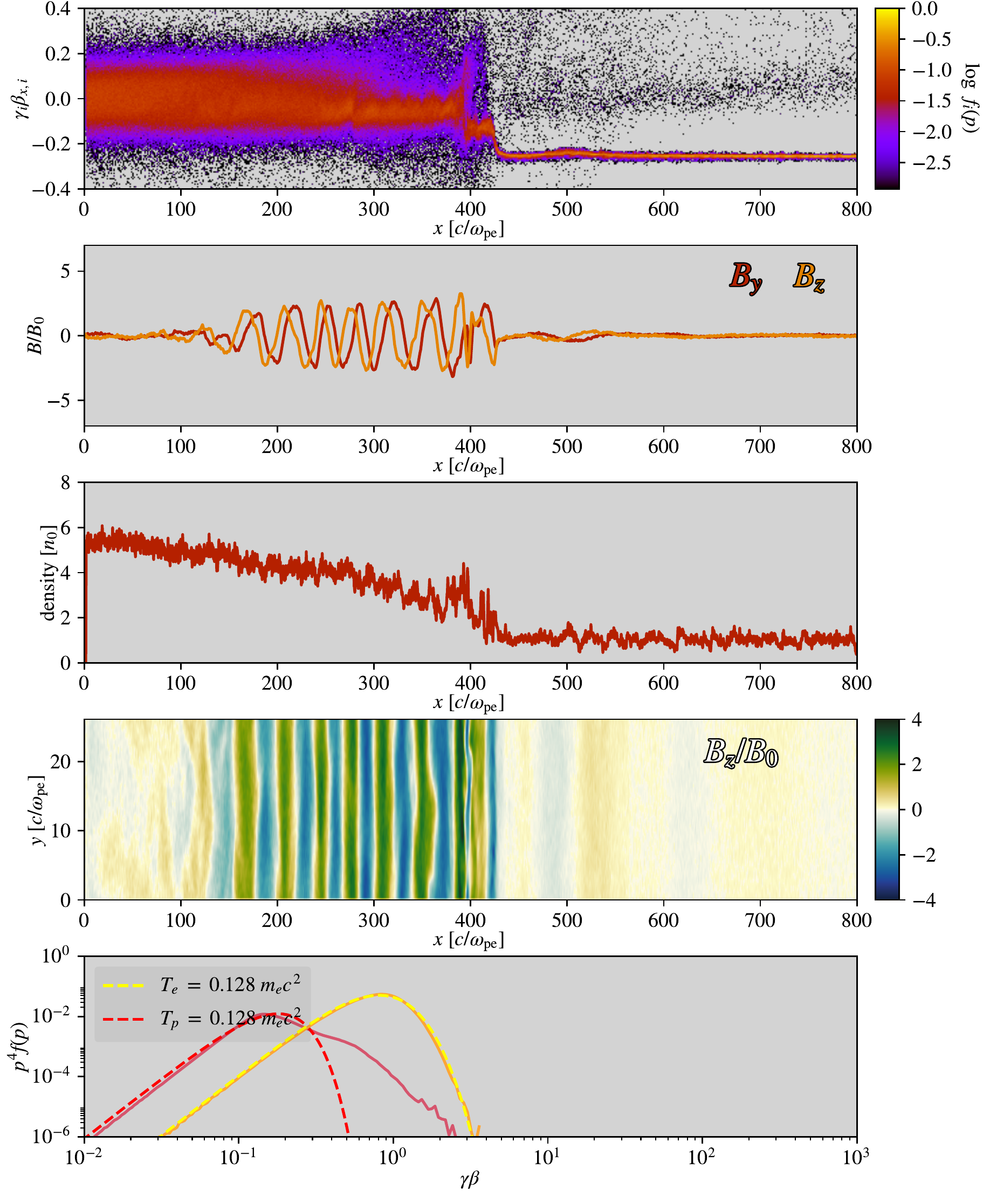}

	\caption{\label{fig:pic1} From top to bottom: $p_x$-$x$ phase space, transversely averaged fields $B_y$ and $B_z$, density profile, $B_z$ field in 2D, and spectra of particles. Two cases are shown for the beam velocity: $v = 0.5~c$ ($M_\text{A} \approx 6$) at time $t = 3500~\omega_{pe}^{-1} = 875~\omega_{pi}^{-1}$ on the left, and $v = 0.25~c$ ($M_\text{A} \approx 3$) at $t = 5400~\omega_{pe}^{-1} = 1350~\omega_{pi}^{-1}$ on the right.}
\end{figure*}

For Mach numbers $\lesssim 5$, the shock ramp forms quite gradually. In Fig.~\ref{fig:pic1} (the case where $M_\text{A} \approx 3$), we see that the upstream wave remains quasi-linear for a longer period of time, which postpones the beginning of shock reformation. There is a peak in the $E_x$ profile presented in Fig.~\ref{fig:pic4}, which indicates the overall slowing down of the plasma flow. For $M_\text{A} \approx 3$, this peak appears at the junction of the upstream and downstream waves, which thus clearly separates the upstream and downstream flows. The resonant wave that triggered this shock did not change its properties significantly while it was entering the non-linear stage. The weak upstream wave was excited later by the escape ions. The downstream structure actually corresponds to the imprint of the resonant wave, which entered the non-linear regime, but is still not affected by the shock reformation. The upstream oscillations in $E_x$ profile for $M_\text{A} \approx 6$ are due to a periodical change of the plasma velocity induced by the upstream wave, and the strong peak is due to a reformed shock.

In the later stages, when the reformation process is stabilized, we observe that the upstream wave changes its wavelength and polarization, while being advected throughout the shock interface. It becomes resonant with the velocity of the upstream flow. As the shock propagates, the wave stays imprinted in the downstream plasma. By using linear equations, we showed that the resonant instability remains static in the downstream frame and saturates when its amplitude reaches the value $\sim B_0$. We also showed that the finely tuned resonant wavelength ($\lambda_\text{res} \approx \pi r_{gi}$) and the left circular polarization of the instability are the properties we get when the two plasmas are subjected to shock-like conditions. The downstream structure thus shows a great similarity with the resonant instability that we studied in Sec.~\ref{sec:resalfinst}. Although, the linear theory does not apply in this non-linear stage, it may point to why the wavelength and polarization change.

\begin{figure}[!ht]
	\centering
	\includegraphics[width=\linewidth,height=2.3cm]{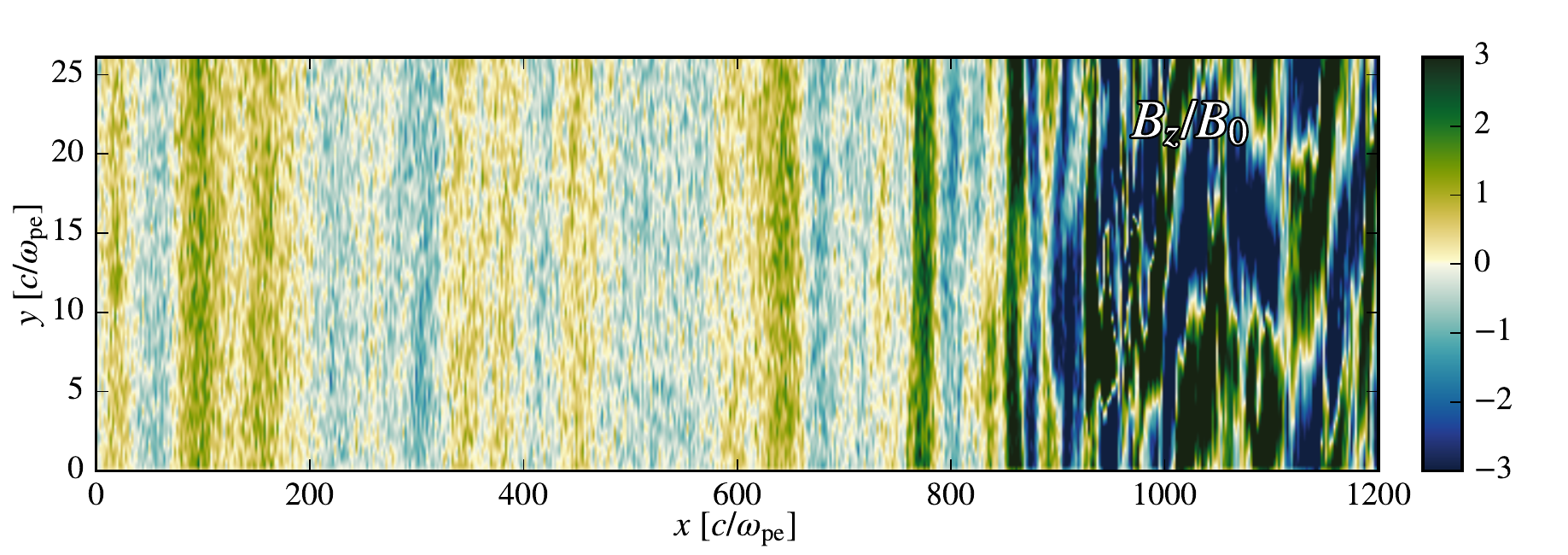}
	\includegraphics[width=\linewidth,height=2.3cm]{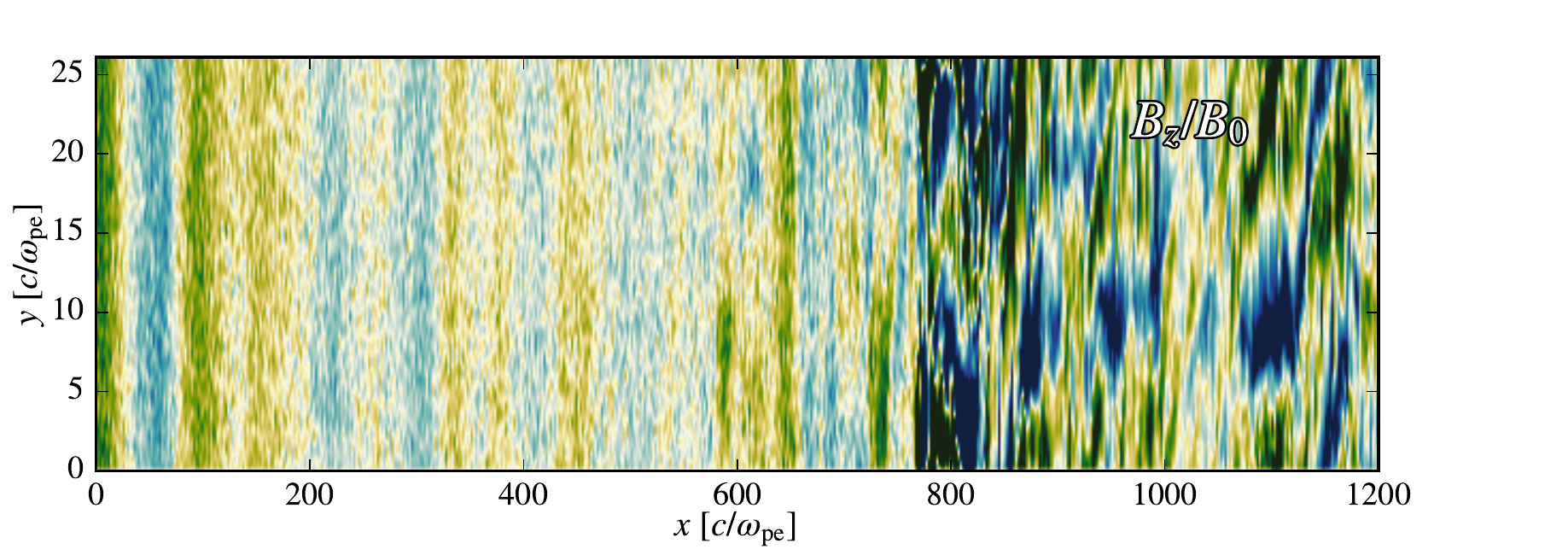}
	\includegraphics[width=\linewidth,height=2.3cm]{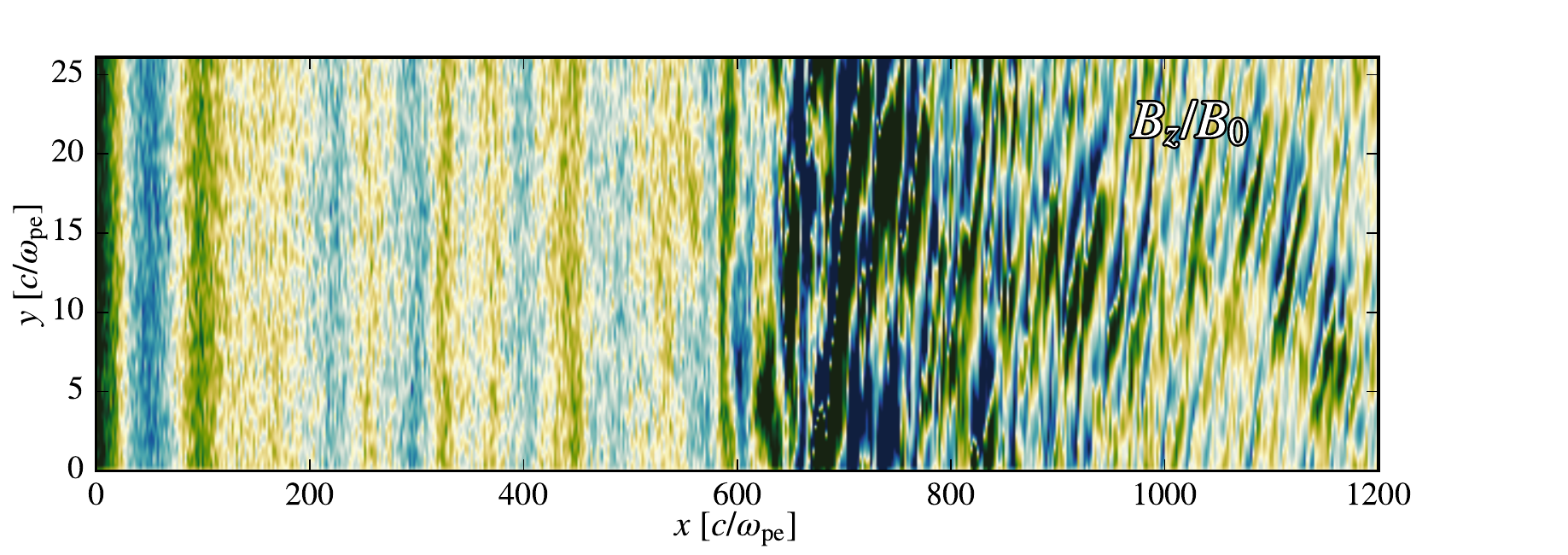}
	\includegraphics[width=\linewidth,height=2.3cm]{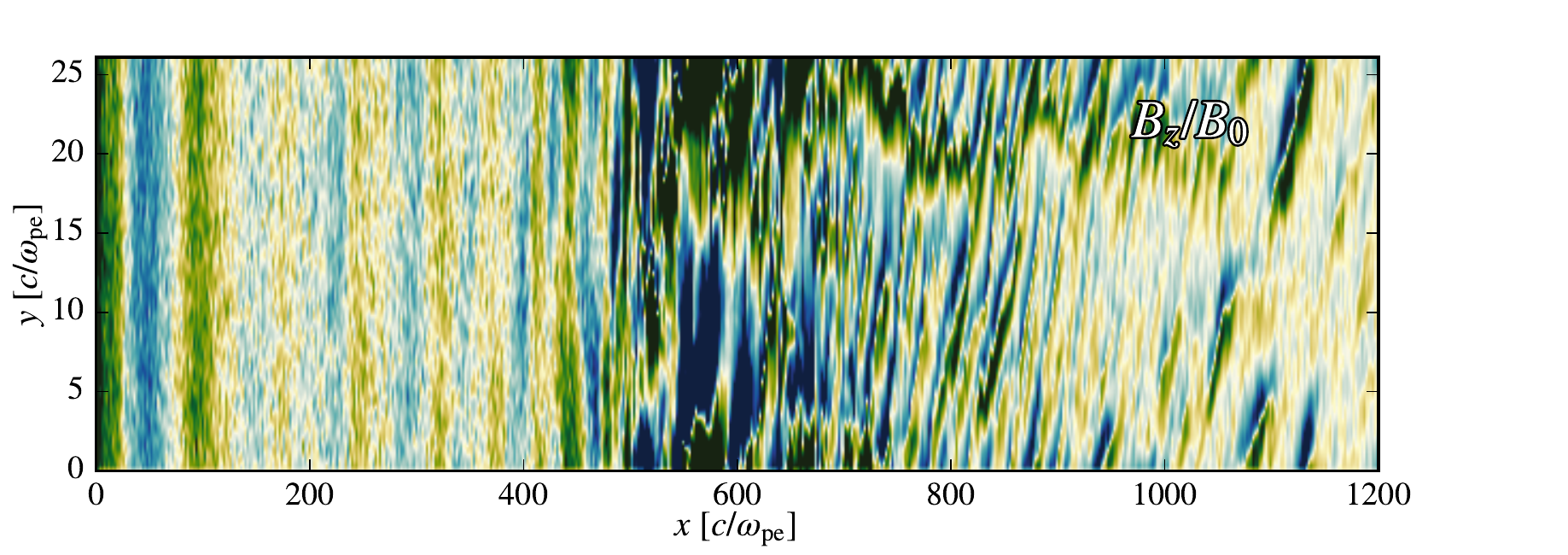}
	\includegraphics[width=\linewidth,height=2.3cm]{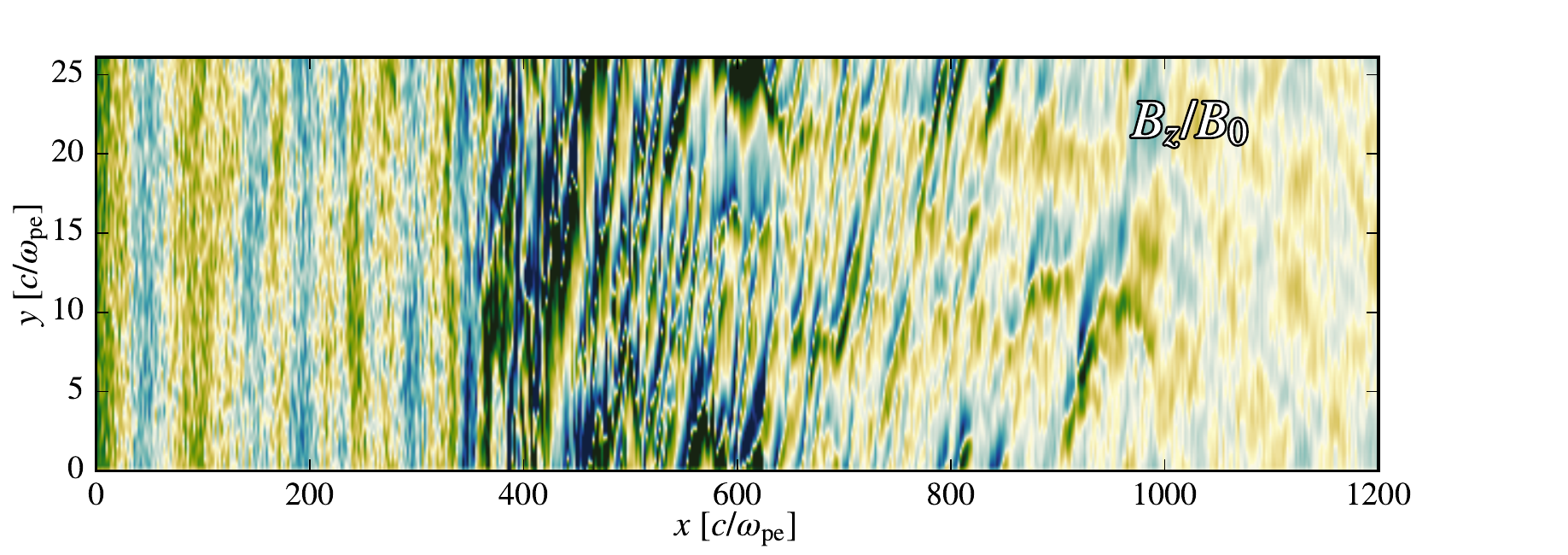}
	
	\caption{\label{fig:pic2} Two-dimensional plot of the transverse magnetic field component $B_z$ in the units normalized to $B_0$, obtained for $\gamma_0 = 15$ at times $t = \{ 1350, 1800, 2250, 2700, 3150 \} \ \omega_{pe}^{-1}$, from bottom to top, respectively. Magnetic field lines are inclined to an angle $\theta = 15^\circ.$}
\end{figure}

\begin{figure*}[!ht]
	\centering
	\includegraphics[width=0.49\linewidth]{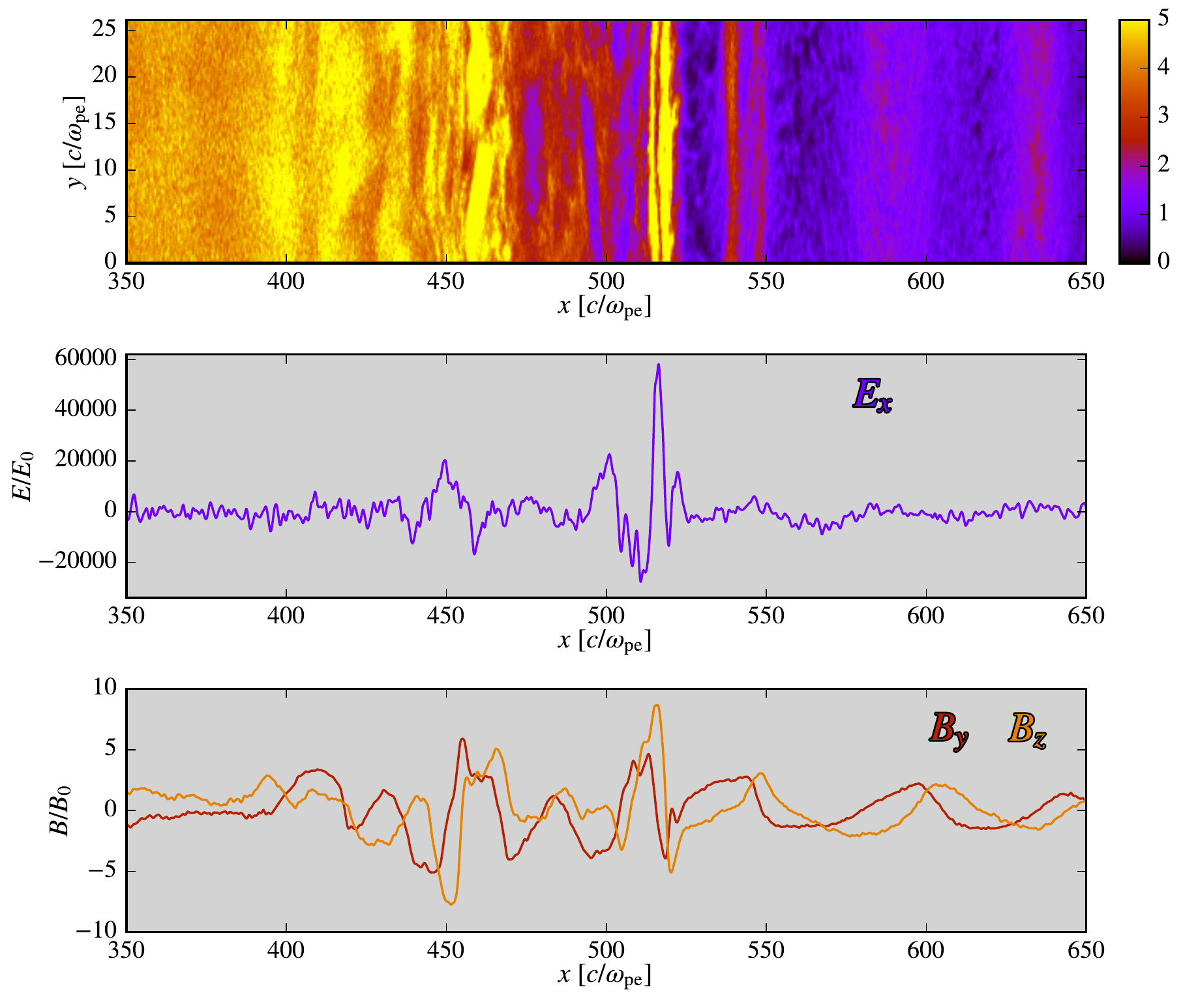}
	\includegraphics[width=0.49\linewidth]{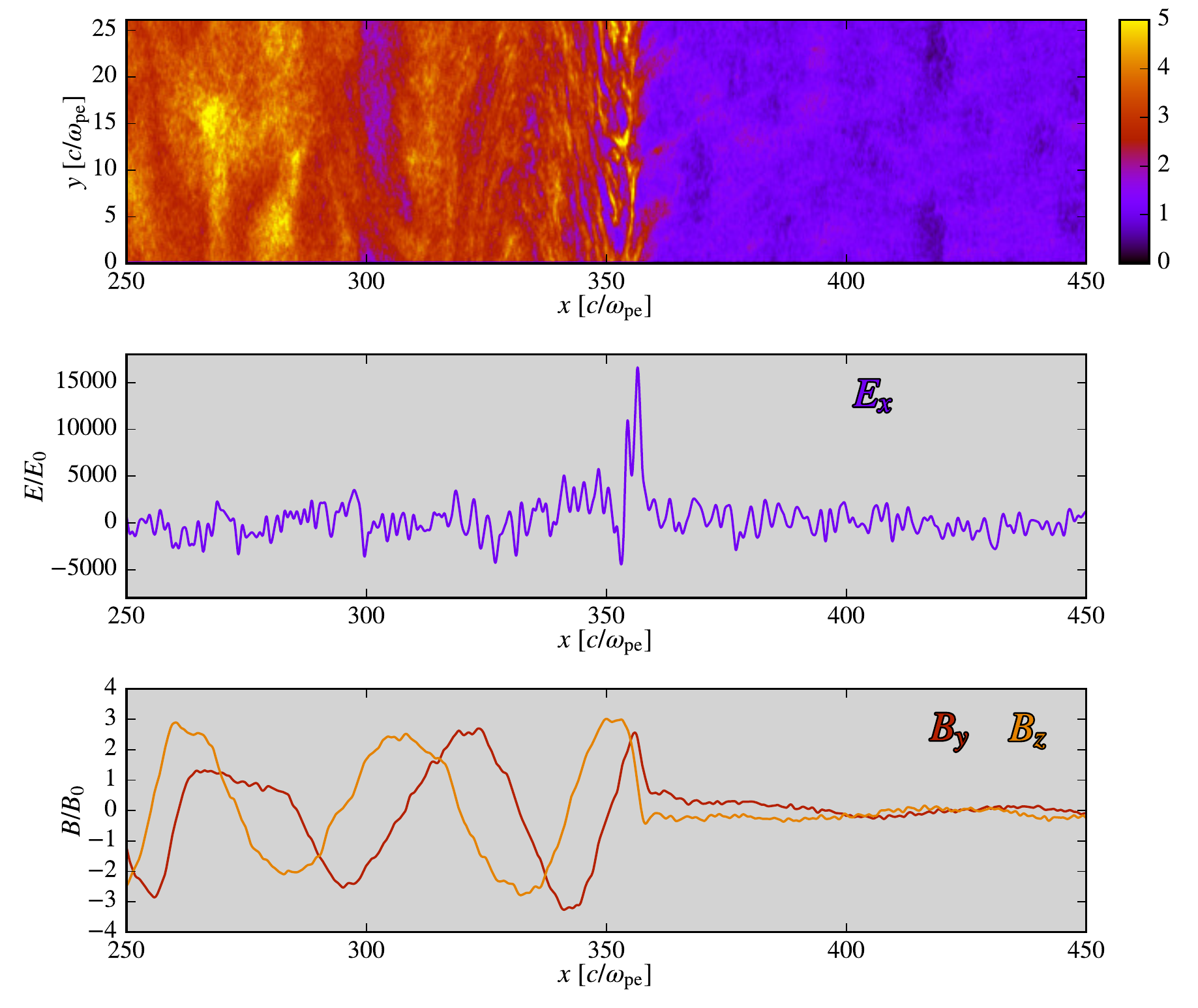}
		
	\caption{\label{fig:pic4} From top to bottom: density field in 2D, transversely averaged fields $E_x$, $B_y$ and $B_z$. The two cases of $M_\text{A} \approx 6$ (left) and $M_\text{A} \approx 3$ (right) are the same as in Fig.~\ref{fig:pic1}. The peaks in $E_x$ profiles are associated with the collective slowing down of the flow: at the first maximum of the downstream instability (right) and at every maximum of the downstream and compressed upstream instabilities (left) of the existing and reformed shocks, respectively.}
\end{figure*}

In the case of an inclined magnetic field (see Fig.~\ref{fig:pic2}), we see that the wavevector of the upstream mode is aligned with the lines of the inclined magnetic field. The resonant wave, however, propagates in the direction of the flow. If we consider the case of super-Alfv\'enic flow with a non-relativistic velocity ($c \gg v_0 \gg v_\text{A}$), we find from the Eq.~(\ref{eq:rlinstii3}) that the approximate analytical expression for the growth rate in a close region around the maximum is:

\begin{equation}
	\gamma_\text{M}^2 = \frac{\eta}{\eta + 1} \kappa^2 - 2 \kappa + 1,
	\label{eq:yk}
\end{equation}

\noindent where $\gamma_\text{M} = \gamma / \omega_{ci}$ is the relative growth rate, and $\kappa = {\bf k} \cdot {\bf v}_0 / \omega_{ci}$ is the relative wavenumber in the direction of the flow ($\eta$ is a density ratio, as before). Since the relative wavenumber is defined by the scalar product ${\bf k} \cdot {\bf v}_0$, it will be the highest for the wave modes that are aligned with the direction of the flow. Therefore, from Eq.~(\ref{eq:yk}), we find that the resonant mode, which propagates in that direction, will grow at the fastest rate. Consequently, this means that the wavevector of this mode should be aligned with the flow, rather than being aligned with ${\bf B}_0$. We observe such behavior of the resonant wave in the simulations. It is observed during the linear stage (before a shock is triggered) as well as in the non-linear stage, when shock reformation is ongoing.

In the case of $M_\text{A} \approx 3$ (see Fig.~\ref{fig:pic1}), we find that the reason the density ramp behind the shock is gradual may be a very low Afv\'enic Mach number. On the $p_x-x$ diagram, it can be seen that the flow is dispersed and compressed by a factor $\sim 3$ in the region where the instability prevails. The plasma continues to drift along ${\bf B}_0$ with the mean velocity of $\sim -0.1~c$. Once the plasma reaches the wall, it reflects with a velocity of $\sim 0.1~c$. It thus doubles its density up to a ratio $\sim 6$. Because the two counter-streaming plasmas are sub-Alfv\'enic in the downstream, the wave mode there changes. The wave decreases its amplitude and allows the plasma to diffuse freely towards the upstream. This region spreads between the wall on the left and the point on the right, where the compression is $r \sim 4$. It gradually expands across the downstream, creating the observed density profile. We observe this during the whole simulation run.

\subsection{\label{sec:picu}Upstream non-resonant instability and shock reformation}

In the simulations, we observe that the ions not captured by the resonant wave during the shock triggering, escape and form the initial pulse of the return current. We find that this current excites a circularly polarized EM wave in the upstream, which is the first to trigger shock reformation (once it grows to the amplitude of $\sim 4~B_0$). The initial bunch of escape ions continues to propagate and pre-heats the plasma ahead of the shock. It thus forms a weak precursor in the upstream. In the next stage, shock reformation completely overtakes the process of particle scattering (thermalization) and further mediates the shock~\cite{sda}. The return current is formed by specular reflection of ions from the reforming shock barrier, as described in the minimal model for ion injection~\cite{numrefprtl}. The upstream wave is then coupled with the current of reflected ions to create a self-sustained structure. Depending on the effective Mach number, the upstream mode can be both -- resonant or non-resonant~\cite{amatoblasi}, where the latter is also known as Bell's instability~\cite{NR}. We also find that the weak-beam instability that we studied in Sec.~\ref{sec:resalfinst} has the same property. Lower Mach numbers of the current of reflected ions cause the upstream instability to be resonant, and higher Mach numbers cause it to be non-resonant. As the instability accelerates particles through the process of DSA~\cite{sda}, the return current penetrates deeper into the upstream. By this process, the instability spreads farther upstream, and the particles are accelerated to higher energies. The non-thermal peak appears in the particle spectra, shortly after the shock initially forms and the return current initiates the upstream mode.

In Fig.~\ref{fig:pic1} (the left column), we observe the upstream instability as a right circularly polarized wave that is a negative helicity wave. In the region farther away from the reforming shock barrier ($800 - 1300~c/\omega_{pe}$), the wave is of a smaller amplitude ($< B_0$), and it does not perturb the upstream flow, nor the return current. In this region, the upstream wave can approximately be considered like it is in the linear stage. Therefore, we model this wave as a weak-beam, cold plasma instability~\cite{microinst,emibinst} that we also analyzed here in Sec.~\ref{sec:resalfinst}. As Eq.~(\ref{eq:rlinstii3}) implies, the instability is kinematically bound to the upstream plasma because the upstream flow is much denser than the return beam of ions. By its polarization and wavelength, the upstream instability is non-resonant with the upstream flow. However, reflected ions ``feel'' it as a left circularly polarized wave -- it therefore interacts resonantly with the return current. The cyclotron resonant condition $\omega_r \simeq {\bf v}_\text{ri} \cdot {\bf k} - \omega_{ci}$ applies to the return beam, where ${\bf v}_\text{ri}$ is the velocity of returning ions relative to the upstream flow.

Here, we apply the results from Sec.~\ref{sec:resalfinst} to estimate the amount $\eta_\text{ri}$ of reflected ions that drive the weak-beam instability, which has the same properties that we observe for the upstream wave. We assume that ions are reflected from the periodically reforming shock barrier and that the reflection occurs in the downstream frame (as shown in~\cite{numrefprtl}). The mean velocity of injected ions in the ISM (lab) frame is then $v_\text{ri} \sim 3.23~v_\text{sh}$, which corresponds to the injection energy $E_\text{inj} \sim 5~E_\text{sh}$ in the shock frame (which we observe). From Fig.~\ref{fig:pic1}, we see that the wavelength of the upstream mode is $\lambda_\text{up} \approx 50~c / \omega_{pe}$. We inspect Eq.~(\ref{eq:rlinstii3}) to find the beam density $\eta_\text{ri}$, for which the growth rate $\gamma (k)$ has a maximum at

\begin{equation}
k \cdot r_{gi} = k \cdot \frac{v_{ri}}{\omega_{ci}} = \frac{2 \pi}{\lambda_\text{up} / \left[ \frac{c}{\omega_{pe}}\right] }\cdot \frac{v_{ri}}{v_\text{A}} \sqrt{\frac{m_i}{m_e}}. \nonumber
\end{equation}

In the case $M_\text{A} \approx 6.34$ (Fig.~\ref{fig:pic1}), for ions which are injected with a velocity $v_\text{ri}$ and excite a wave with $k \cdot r_{gi} \approx 13.4$, we obtain that $\eta_\text{ri} \approx 0.038$. This result finely agrees with the minimal model~\cite{numrefprtl}, where injection fraction is $\sim 3.6 \%$. However, if we apply the same calculation to the case of a shock presented in the minimal model, where $M_\text{A} = 20$ and $\lambda_\text{up} \sim 20~c / \omega_{pi}$, we find that $\eta_\text{ri} \sim 0.01$. If we use the same duty cycle for ion injection $\sim 25 \%$ (meaning that the probability of a particle being captured by a reforming barrier is $\mathcal{P}\sim 0.75$), as in~\cite{numrefprtl}, we find that the number of SDA cycles needed to obtain $\eta_\text{ri} = (1-\mathcal{P})^\mathcal{N}$ is $\mathcal{N} \sim 3.3$. This implies the injection energy $E_\text{inj} \sim 6~E_\text{sh}$. In our case of $M_\text{A} \approx 6.34$, the number of cycles is $\mathcal{N} \sim 2.4$, the same as in~\cite{numrefprtl}. It means that in the case of a quasi-parallel shock, specularly reflected ions are energized via SDA. Only those ions that experience $\mathcal{N}$ cycles of SDA are able to overcome the shock barrier and drive the upstream instability.

Therefore, we relate the weak-beam, cold plasma instability with the upstream wave, which we observe in the simulations. In the non-linear stage, this instability triggers shock reformation. While being transmitted through the shock, the upstream instability changes its polarization and wavelength to form the (non-linear) resonant wave in the downstream. The amplitude of the instability grows to the value of $(1.5-2.5) B_0$ just in front of the shock interface. There, the upstream flow slows down and becomes compressed. The transverse component of the magnetic field also compresses, and the amplitude of the instability increases to $ r \cdot (1.5-2.5) B_0 \approx (6-10) B_0$, which is commonly observed in kinetic simulations~\cite{sda,rvsnr}. At the shock transition, due to longitudinal compression, the instability wavelength also shrinks by the same ratio.

\begin{figure}[!ht]
	\centering
	\includegraphics[width=\linewidth]{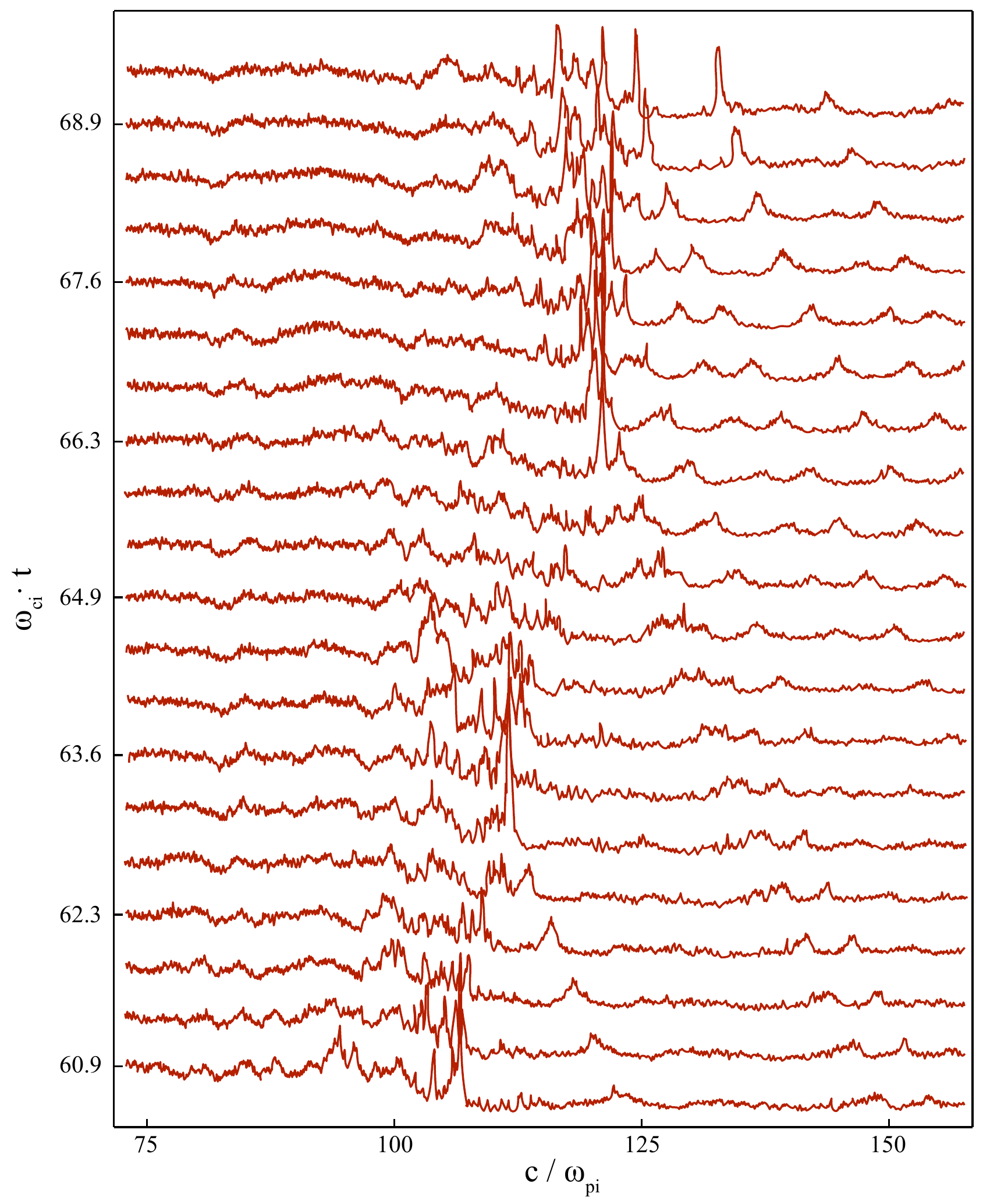}
	
	\caption{\label{fig:pic3} The density profile captured at different times shows an ongoing shock reformation process. On the horizontal axis, the \emph{x}-coordinates are given in units of the ion skin depth ($c / \omega_{pi}$); and on the vertical axis, the time is given normalized to the ion gyro-period $T_{ci} = 2 \pi / \omega_{ci}$.}
\end{figure}

We observe that in the region of the reforming shock barrier, the beam of returning ions is periodically disrupted by the upstream mode in the longitudinal direction. One part of the returning ions slows down and becomes trapped locally along the path, at places where the amplitude of the upstream instability reaches its maximum. We find that this deceleration occurs because of the wave resonance with the velocity of returning ions. Due to a slightly different velocity of the upstream mode with regard to the upstream flow, we also find that the flow is non-resonantly perturbed by the instability. In the region where the wave becomes non-linear, ions fall down into a periodic potential well (that is imposed by the upstream wave), and form clumps. These clumps of ions are bound to the instability, which advects them to the shock. While approaching the shock region, the plasma clumps gradually slow down and become compressed in exactly the same way as the upstream instability. Density spikes with $\eta \sim 1.5$ thus form, and near the shock, they become further compressed to $\eta \sim 6-10$, as we observe in the density profile shown in Fig.~\ref{fig:pic3}. The spikes periodically reform ahead of the shock, at the regions where the growing instability becomes compressed. This is the typical signature of the shock reformation process that is commonly observed in kinetic simulations~\cite{locshckreform,sda,numrefprtl}.

We also notice that the upstream instability may be amplitude modulated, as shown in Fig.~\ref{fig:pic1}. This modulation directly reflects the upstream plasma density by making the particles bunch more or less in synchronicity with the change of the wave envelope. We observe this in the density profiles shown in Fig.~\ref{fig:pic3}, where we found that the amplitude of the spikes is modulated in the same way as the upstream instability. This means that the shock might reform with the periodicity of the carrier wave (the upstream instability), which we see in the upper-half of the density graph. However, in the lower-half of the graph, we observe the regions where spikes are of a low amplitude or absent. These regions correspond to the minima in the amplitude of the modulation waveform. Contrary to them, the regions with amplified spikes correspond to the maximum amplitude of the modulating wave. As a result of the collective effect, within the observed time interval, the shock reforms on two time scales: shorter -- with the length of one oscillation of upstream instability, and longer -- with the length similar to the wavelength of the downstream mode. There is a shock-pause in profiles that reside in the middle of Fig.~\ref{fig:pic3} (around $\omega_{ci} t \sim 65$). This is clearly a consequence of shock reformation on a larger time scale, which appears due to the modulation of the whole process.

We now give a brief overview of the results. In the case of a parallel shock, which we presented here, we found that the shock reform periodicity coincides with the upstream wave periodicity. We related this wave by its properties to the weak-beam instability. We showed that the instability disrupts the return current resonantly, and perturbs and decelerates the flow non-resonantly. The deceleration of the plasma is large enough to increase the flow density and, thereby, compress the field and reform the shock.

\section{\label{sec:results}Results and discussion}

We analyzed the interaction of the two interpenetrating, cold, quasi-neutral and magnetized ion-electron plasmas of comparable densities. We considered the plasma flowing quasi-parallel to the lines of the background magnetic field $B_0$, with a constant velocity $v_0$ through another static plasma. The linear fluid equations showed that, if $v_0$ becomes super-Alfv\'enic, the parallel R and L waves as well as the perpendicular X and O waves all become unstable. We showed that the instability grows only within a certain range of wavelengths, where the growth rate has a maximum. For comparable densities of the two plasmas, this maximum resides in the range of the resonant wavelengths. To understand what happens in the first stages of the non-linear regime, we started with the case of two colliding plasmas that have equal densities. We combined the linear theory with the results of the test particle simulations~\cite{rpbsw} and we found that due to resonance, the instability (which grows to the amplitude $\sim B_0$) becomes able to capture the ions on its propagation path and, thus, trigger shock formation.

We ran PIC simulations of quasi-parallel (sub)relativistic collisionless shocks, considering different ion masses and shock velocities. At the very first phases of a shock formation, we detected a resonant wave of the amplitude $\sim B_0$ and a wavelength $\sim \lambda_\text{res}$. As we explained in Sec.~\ref{sec:pic}, this wave largely matches the resonant instability that we studied in Sec.~\ref{sec:resalfinst}. During this very short transient, we observed that ions are strongly scattered by the wave. Therefore, it seems that initially the resonant instability disrupts the flow and triggers thermalization, and then shortly afterwards, a shock is formed in the non-linear stage.

Ions initially not captured by the resonant mode (those that reside in part of the flow which is not encompassed by the grown instability), thus manage to escape and form the initial return current. This sparse bunch of ion plasma excites a wave of opposite polarization ahead of the forming shock. Because this wave quickly grows to a non-linear instability that advects toward the shock, it is the first one to trigger the shock reformation process. Reformation quickly takes over, and it further balances the coupling between the upstream wave and the return current. Polarization of the upstream wave (which is commonly observed in kinetic simulations) is such that it is non-resonant with respect to the upstream flow. It is known as the cosmic-ray-induced streaming instability~\cite{amatoblasi}, whose non-resonant mode is also referred to as Bell's instability~\cite{NR,sda}, or hybrid non-resonant instability~\cite{rvsnr}. At the same time, we find that polarization of this wave is resonant with respect to the gyro-motion of the outflowing ions. Based on this, we related the upstream wave to the R-mode micro-instability (positive helicity whistler wave) that is studied here in the weak-beam case and is already known as \emph{EM ion/ion right-hand resonant} instability~\cite{emiiinst,microinst}.

In Sec.~\ref{sec:pic}, we explained how this self-sustaining instability governs the shock reformation process in the case of a quasi-parallel shock. The inflowing upstream plasma is non-resonantly perturbed by the instability, and reformation is triggered in the region where the wave grows to a few $B_0$ in amplitude. The return beam of ions becomes periodically disrupted and slowed down by the resonant mechanism. Nevertheless, one part of the returning ions continues to propagate, thus expanding the acceleration region for the DSA mechanism. We used the properties of the upstream wave that we observed at a later stage, when the wave-beam coupling is stabilized. In Sec.~\ref{sec:pic}, we showed that in the region further away from the shock, the weak-beam case~\cite{microinst,emibinst} can be applied to the return current. In the calculation explained in Sec.~\ref{sec:picu}, where we used the linear theory presented in Sec.~\ref{sec:resalfinst}, we obtained the value of $\sim 3.8 \%$ for ion injection. For comparison, in the minimal model~\cite{numrefprtl}, the fraction of injected ions is found to be $\sim 3.6 \%$ of the inflowing ions.

The wave observed at the shock is due to advection and compression of the upstream instability. The shock is reformed at the first peak of the compressed upstream wave. Afterwards, the wave slows down and couples to a downstream structure, which is of opposite polarization and of a significantly different wavelength. We found that the wavelength of the downstream instability is highly resonant ($\sim \pi r_{gi}$). In Sec.~\ref{sec:resalfinst}, using linear equations we related this resonance to the case of flow compression of $\sim 4$. Therefore, the compression ratio similar to the one from Rankine-Hugoniot conditions can also arise due to resonant wave-plasma coupling during shock (re)formation. Nevertheless, because the shock is heavily non-linear, the results of the linear theory cannot be applied with complete certainty. On the other hand, we found that for Mach numbers $\lesssim 5$, the shock forms quite gradually. The upstream wave remains quasi-linear for a longer period of time, which postpones the beginning of shock reformation. We found that the instability, which triggers the initial particle scattering, does not change its waveform throughout shock formation. The peak in $E_x$ profile (which is clearly associated with the overall slowing down of the flow), and the boost in the wave phase (which also appears at the same point), thus clearly distinguish the upstream and the downstream wave in the stage before the beginning of shock reformation.

As in~\cite{numrefprtl}, we observe that shock reformation happens on the scales of the instability wavelength. However, we showed that the upstream wave is amplitude modulated. The envelope of the modulated wave changes with the waveform of the downstream mode. Therefore, we think it is very likely that the downstream instability modulates the intensity of the return current, which in return modulates the amplitude of the upstream mode. Consequently, this leads to the amplification or suppression of the whole process on a larger scale, as the envelope increases or decreases with the waveform of the downstream instability.

The composite picture we get here, shows certain similarities to the model presented in~\cite{Malkalfv}. Firstly, the Alfv\'enic type wave is excited in the upstream. Secondly, the instability constrains the escape of ions by resonantly trapping and advecting them back to the shock. However, perturbation of the upstream flow by the wave and, therefore, the shock reformation process are not accounted therein. Kinetic simulations~\cite{sda,nrelshocks} show that a return current is formed by the population of reflected ions that escape the reforming barrier. Even more important is that the major ion population cannot thermally leak~\cite{noleak,numrefprtl}, not even from the near downstream.  The more energetic, heavier nuclei can, however, diffuse from the \emph{far downstream} and become re-accelerated while crossing the shock, as explained in~\cite{espresso,uheacc,DSRA}.

\section{\label{sec:summ}Summary}

In addition to the existing models, in this paper we want to disclose the significance of resonant micro-instabilities and their possible role in the triggering, transmission and reformation of a shock wave. We give a summary of our results found in the theory and observed in PIC simulations:

1. From the linear theory, we found that resonant circularly polarized micro-instabilities can be driven by two interpenetrating, cold, quasi-neutral and magnetized ion-electron plasmas of comparable densities.

2. During the very first phases of a collisionless shock formation, we found that a resonant wave is excited in PIC simulations. By its properties, we related this wave to resonant micro-instability, which is driven by the flow itself (a strong beam). Because of resonance, we found that the wave strongly scatters the particles and triggers thermalization. Shortly afterwards, shock conditions are formed by non-linear processes.

3. Ions initially not captured by a resonant wave, escape and form the initial return current. This current excites the wave ahead of the forming shock, which is the first to trigger shock reformation process. We modeled the wave by the same type of micro-instability that we found in the weak-beam case. The amount of injected ions that we got from the equations finely agrees with the minimal model~\cite{numrefprtl} for ion injection from quasi-periodic shock barrier.

4. The downstream wave is likely due to the advected upstream wave, which changes its wavelength and polarization at the shock interface and becomes resonant with the upstream flow. The linear theory does not apply because of the high non-linearity of the wave. Still, our equations indicate that the flow compression of $\sim 4$ can naturally arise as a consequence of the highly resonant micro-physical process.

5. We found that the upstream wave is amplitude modulated by the return current. The modulation leads to the amplification or suppression of the reformation process on a larger scale, as the envelope increases or decreases with the waveform of the downstream mode. Consequently, two characteristic shock reformation scales emerge.

In view of the points outlined here, we now give the answer to the entitled question. Resonant micro-instabilities can indeed trigger the formation of quasi-parallel magnetized collisionless shocks. However, their role is constrained to a very short transient stage, until the reformation process begins. This strong-beam instability remains resonant even for higher Alfv\'enic Mach numbers, which makes it a reliable candidate for shock triggering. The same type of micro-instability, but with opposite polarization in the upstream, then reforms and transmits the shock. Downstream structure is likely to appear as a consequence of the non-linear mode change of the upstream instability.

In the end, it seems like the whole shock structure develops as the particles are accelerated to higher energies. Quasi-parallel magnetized collisionless shocks therefore represent highly complex self-sustaining systems, which can provide us with detailed insight into the shock micro-physics and physics of particle acceleration to cosmic-ray energies.

\acknowledgments
Gratitude for this work I owe to my advisor Bojan Arbutina. For valuable discussions and suggestions, I also thank Dejan Uro{\v s}evi{\'c} and Marko Pavlovi{\'c}. For her great help in preparing this paper, I thank Aleksandra {\'C}iprijanovi{\'c}. Special thanks to both reviewers for their very constructive comments. PIC simulations were run on PARADOX-IV supercomputing facility at Scientific Computing Laboratory of the Institute of Physics Belgrade, and also on cluster ``Jason'' of Automated Reasoning Group (ARGO) at the Department of Computer Science, Faculty of Mathematics, University of Belgrade. The results of PIC simulations were visualized by ``Iseult'' - a GUI written by Patrick Crumley. This research has been supported by the Ministry of Education, Science and Technological Development of the Republic of Serbia under project No. 176005.

\appendix

\section{\label{sec:appequations}Equations of cold interpenetrating plasmas}

In the regime of small oscillations, the first order perturbation equations for the flowing plasma in the frequency domain become:

\begin{eqnarray}
	\left( -i\omega + i{\bf k} \cdot {\bf v}_0 \right) {\bf v}_i && = \frac{q_i}{m_i} \left( {\bf E}_1 + {\bf v}_i \times {\bf B}_0 + {\bf v}_0 \times {\bf B}_1 \right),
	\label{eq:lfi}
	\\
	\left( -i\omega + i{\bf k} \cdot {\bf v}_0 \right) {\bf v}_e && = \frac{q_e}{m_e} \left( {\bf E}_1 + {\bf v}_e \times {\bf B}_0 + {\bf v}_0 \times {\bf B}_1 \right).
	\label{eq:lfe}
\end{eqnarray}

\noindent In the zeroth order, Eqs.~(\ref{eq:nli})--(\ref{eq:nle}) relate the equilibrium quantities as~\cite{collplasmastroph}

\begin{equation}
	0 =  {\bf E}_0 + {\bf v}_0 \times {\bf B}_0 ,
	\label{eq:lf0}
\end{equation}

\noindent and show us that the flow of particles through the constant magnetic field induces a constant electric field, which then opposes the magnetic force imposed on the moving charges. The flow remains undisturbed and the particles continue to drift through the background magnetic field.

For the stationary plasma (${\bf v}_0 = 0$), Eqs.~(\ref{eq:lfi}) and (\ref{eq:lfe}) become:

\begin{eqnarray}
	-i\omega {\bf v}_i = \frac{q_i}{m_i} \left( {\bf E}_1 + {\bf v}_i \times {\bf B}_0 \right),
	\label{eq:lsi}
	\\
	-i\omega {\bf v}_e = \frac{q_e}{m_e} \left( {\bf E}_1 + {\bf v}_e \times {\bf B}_0 \right).
	\label{eq:lse}
\end{eqnarray}

For simplicity, from now on, index ``1'' is dropped from the symbols for variable EM field components ${\bf E}_1$ and ${\bf B}_1$.

Perturbation velocities of stationary plasma particles ${\bf v}_{i,e}^s$ are obtained from Eqs.~(\ref{eq:lsi})--(\ref{eq:lse}) and can be expressed as functions of fields:

\begin{equation}
	{\bf v}_{i,e}^s = \frac{q_{i,e}}{m_{i,e}}
		\begin{pmatrix}
			-\dfrac{1}{i \omega} & 0 & 0 \\
			0 & \dfrac{i \omega}{\omega^2 - \omega_{ci,e}^2} & \dfrac{- \omega_{ci,e}}{\omega^2 - \omega_{ci,e}^2} \\
			0 & \dfrac{\omega_{ci,e}}{\omega^2 - \omega_{ci,e}^2} & \dfrac{i \omega}{\omega^2 - \omega_{ci,e}^2} \\
		\end{pmatrix}
		\cdot {\bf E},
	\label{eq:vies}
\end{equation}

\noindent where cyclotron frequency of the ions/electrons is given by $\omega_{ci,e} = q_{i,e} B_0 / m_{i,e}$.

Similarly, the perturbation velocities ${\bf v}_{i,e}^f$ of the flowing plasma particles are obtained from Eqs.~(\ref{eq:lfi})--(\ref{eq:lfe}). By using Faraday's law of induction $\omega {\bf B} = {\bf k} \times {\bf E}$, the component of these equations, which contains a perturbed magnetic field, can be written as

\begin{eqnarray}
	{\bf v}_0 \times {\bf B}~&&= -\frac{({\bf v}_0\cdot{\bf k})}{\omega}~{\bf E} + \frac{({\bf v}_0\cdot{\bf E})}{\omega}~{\bf k} \nonumber
	\\
	&&= -\frac{({\bf v}_0\cdot{\bf k})}{\omega}~{\bf E} + \frac{{\bf k} {\bf v}_{0}^{\text{T}}}{\omega} {\bf E}.
\end{eqnarray}

By substituting this into the equations of the flowing plasma, the perturbation velocities are found to be 

\begin{eqnarray}
	{\bf v}_{i,e}^f &&= \frac{q_{i,e}}{m_{i,e}}
		\begin{pmatrix}
			-\dfrac{1}{i \xi \omega} & 0 & 0 \\
			0 & \dfrac{i \xi \omega}{\xi^2 \omega^2 - \omega_{ci,e}^2} & \dfrac{- \omega_{ci,e} }{\xi^2 \omega^2 - \omega_{ci,e}^2} \\
			0 & \dfrac{\omega_{ci,e}}{\xi^2 \omega^2 - \omega_{ci,e}^2} & \dfrac{i \xi \omega}{\xi^2 \omega^2 - \omega_{ci,e}^2} \\
		\end{pmatrix} \cdot \nonumber
	\\
	&&\cdot \left( \xi + \frac{{\bf k} {\bf v}_{0}^{\text{T}}}{\omega} \right) {\bf E},
	\label{eq:vief}
\end{eqnarray}

\noindent where the term that contains the tensor product is defined by

\begin{equation}
	{\bf k} {\bf v}_{0}^{\text{T}} = 
		\begin{pmatrix}
			k_x v_0^x & k_x v_0^y & k_x v_0^z \\
			k_y v_0^x & k_y v_0^y & k_y v_0^z \\
			k_z v_0^x & k_z v_0^y & k_z v_0^z \\
		\end{pmatrix}.
\end{equation}

\noindent The frequency of the wave is now modified by the collective motion of plasma particles, where modification is made through the parameter

\begin{equation}
	\xi = 1 - \frac{ {\bf v}_0\cdot{\bf k} }{ \omega }.
	\label{eq:ksi}
\end{equation}

The Current is then found by summing up the contributions from all of the species of the flowing and stationary plasmas:

\begin{eqnarray}
	{\bf j} = \sum_{\alpha = i,e}^{\beta = f,s} n_\alpha^\beta q_\alpha {\bf v}_\alpha^\beta + \sum_{\alpha = i,e} n^{f}_{1 \alpha} q_\alpha {\bf v}_0,
	\label{eq:j}
\end{eqnarray}

\noindent where $n^{f}_{1 i,e}$ stands for the density perturbation of the different species of the flowing plasma. It satisfies the linearized continuity equation

\begin{eqnarray}
	- i \xi \omega n^{f}_{1 i,e} + i n_{i,e} {\bf k}\cdot{\bf v}_{i,e}^f = 0,\ \ n^{f}_{1 i,e} = \frac{n^{f}_{i,e}}{\xi \omega}{\bf k}\cdot{\bf v}_{i,e}^f. \nonumber
\end{eqnarray}

Using standard procedure~\cite{intro2plasma_sla} for obtaining the dispersion relation, Amp\`ere and Faraday laws are combined to get a linearized equation of the plasma-wave coupling

\begin{equation}
	{\bf k} \times ({\bf k} \times {\bf E}) + \frac{\omega^2}{c^2} {\bf E} + i \omega \mu_0 {\bf j} = 0.
	\label{eq:afl}
\end{equation}

After substituting Eqs.~(\ref{eq:vies}) and (\ref{eq:vief}) for velocities in Eq.~(\ref{eq:j}), and by using that current, Eq.~(\ref{eq:afl}) can be rewritten as 

\begin{eqnarray}
	{\bf k} \times ({\bf k} \times {\bf E}) &&+ \frac{\omega^2}{c^2} K \cdot {\bf E} = 0.
	\label{eq:aflk}
\end{eqnarray}

\noindent Polarization matrix $K$ can further be separated into two matrices:

\begin{equation}
	K = \kappa + \frac{i}{\epsilon_0 \omega} \sigma_f.
	\label{eq:polm}
\end{equation}

\noindent The first matrix has the form as in the flowless case:

\begin{equation}
	\kappa = \begin{pmatrix}
				P & 0 & 0 \\
				0 & S & -i D \\
				0 & i D & S \\
			 \end{pmatrix},
	\label{eq:polmk}
\end{equation}

\noindent but its parameters are modified by the flow through $\xi$:

\begin{eqnarray}
	P &&= 1 - \sum_{\alpha = i,e}^{\beta = f,s} \frac{\omega_{p\alpha\beta}^2}{\omega^2},
	\label{eq:polmkp}
	\\
	S &&= 1 - \sum_{\alpha = i,e} \frac{\xi^2 \omega_{p\alpha f}^2}{\xi^2 \omega^2 - \omega_{c\alpha}^2} + \frac{\omega_{p\alpha s}^2}{\omega^2 - \omega_{c\alpha}^2},
	\label{eq:polmks}
	\\ 
	D &&= \sum_{\alpha = i,e} \frac{\omega_{p\alpha f}^2}{\omega}\frac{\xi \omega_{c\alpha}}{\xi^2 \omega^2 - \omega_{c\alpha}^2} + \frac{\omega_{p\alpha s}^2}{\omega}\frac{\omega_{c\alpha}}{\omega^2 - \omega_{c\alpha}^2}.
	\label{eq:polmkd}
\end{eqnarray}

\noindent The second matrix fully depends on ${\bf v}_0$ and vanishes completely if the flow speed is zero:

\begin{eqnarray}
	&&\sigma_f = \epsilon_0 \sum_{\alpha = i,e} \omega_{p\alpha f}^2 \cdot \nonumber
	\\
	&&\left( 1 + \frac{{\bf v}_0 {\bf k}^{\text{T}}}{\xi \omega} \right)
	\begin{pmatrix}
			-\dfrac{1}{i \xi \omega} & 0 & 0 \\
			0 & \dfrac{i \xi \omega}{\xi^2 \omega^2 - \omega_{c\alpha}^2} & \dfrac{-\omega_{c\alpha}}{\xi^2 \omega^2 - \omega_{c\alpha}^2} \\
			0 & \dfrac{\omega_{c\alpha}}{\xi^2 \omega^2 - \omega_{c\alpha}^2} & \dfrac{i \xi \omega}{\xi^2 \omega^2 - \omega_{c\alpha}^2} \\
	\end{pmatrix}
	\frac{{\bf k} {\bf v}_{0}^{\text{T}}}{\omega} + \nonumber
	\\
	&&+ \frac{{\bf v}_0 {\bf k}^{\text{T}}}{\omega}
	\begin{pmatrix}
			-\dfrac{1}{i \xi \omega} & 0 & 0 \\
			0 & \dfrac{i \xi \omega}{\xi^2 \omega^2 - \omega_{c\alpha}^2} & \dfrac{-\omega_{c\alpha}}{\xi^2 \omega^2 - \omega_{c\alpha}^2} \\
			0 & \dfrac{\omega_{c\alpha}}{\xi^2 \omega^2 - \omega_{c\alpha}^2} & \dfrac{i \xi \omega}{\xi^2 \omega^2 - \omega_{c\alpha}^2} \\
	\end{pmatrix}.
	\label{eq:polms}
\end{eqnarray}

Taking the wavevector to be in the $x-z$ plane, the dispersion matrix that constitutes a system of equations given by (\ref{eq:aflk}) is

\begin{eqnarray}
	\mathcal{D} &&=
	\begin{pmatrix}
		P - n^2 \sin^2 \theta & 0 & n^2 \sin\theta \cos\theta \\
		0 & S - n^2 & -i D \\
		n^2 \sin\theta \cos\theta & i D & S - n^2 \cos^2 \theta\\
	\end{pmatrix} + \nonumber
	\\
	&&+ \frac{i}{\epsilon_0 \omega} \sigma_f.
	\label{eq:dm}
\end{eqnarray}

\section{\label{sec:appparallelinst}Instability of R and L waves}

To derive $\omega_r$ and $\gamma$, the two frequency ranges are considered: cases when $\xi \omega \sim \omega_{ci} \ll \omega_{ce}$ and $\omega_{ci} \ll \omega_{ce} \sim \xi \omega$.

Eq.~(\ref{eq:drpnormrl}) is at first rearranged as

\begin{eqnarray}
	R (L) = ~&&1 - \frac{\eta~ \xi}{\omega} \cdot \left[ \frac{\omega_{pis}^2}{\xi \omega \pm \omega_{ci}} + \frac{\omega_{pes}^2}{\xi \omega \pm \omega_{ce}} \right] \nonumber
	\\
	&&- \frac{1}{\omega} \cdot \left[ \frac{\omega_{pis}^2}{\omega \pm \omega_{ci}} + \frac{\omega_{pes}^2}{\omega \pm \omega_{ce}} \right].
	\label{eq:rlinst}
\end{eqnarray}

\noindent It is then approximated by the use of previous conditions for the ion-cyclotron and electron-cyclotron frequencies.

\subsection{\label{sec:appparallelinstii}Ion-cyclotron frequencies}

Condition $\xi \omega \sim \omega_{ci} \ll \omega_{ce}$ implies the interaction between ions of the flowing and stationary plasmas. Eq.~(\ref{eq:rlinst}) is therefore approximated by

\begin{eqnarray}
	R (L) \approx ~&&1 - \frac{\eta~ \xi}{\omega} \cdot \left[ \frac{\omega_{pis}^2}{\xi \omega \pm \omega_{ci}} \pm \frac{\omega_{pes}^2}{\omega_{ce}} \right] \nonumber
	\\
	&&- \frac{1}{\omega} \cdot \left[ \frac{\omega_{pis}^2}{\omega \pm \omega_{ci}} \pm \frac{\omega_{pes}^2}{\omega_{ce}} \right].
	\label{eq:rlinstii}
\end{eqnarray}

\noindent Equality $\omega_{pes}^2 / \omega_{ce} = -\omega_{pis}^2 / \omega_{ci}$ is then used to get:

\begin{equation}
	R (L) \approx 1 + \omega_{pis}^2 \left[ \frac{\eta~ \xi^2}{\omega_{ci} (\omega_{ci} \pm \xi \omega)} + \frac{1}{\omega_{ci} (\omega_{ci} \pm \omega)} \right],
	\label{eq:rlinstii2}
\end{equation}

\noindent where $\xi$ is given by Eq.~(\ref{eq:ksi}).

This is the cold plasma relation as given in~\cite{emiiinst,microinst}. In the range of frequencies much lower than $\omega_{ci}$, these two circularly polarized modes merge to form an Alfv\'en wave that is modified by the flow. Therefore, in approximation $\omega \ll \omega_{ci}$, Eq.~(\ref{eq:rlinstii2}) reduces to

\begin{equation}
	n^2 \approx 1 + \frac{c^2}{v_{\text{A}}^2} \left[ 1 + \eta \left( 1 - \dfrac{v_0}{c} n \right)^2 \right],
	\label{eq:rlmerge}
\end{equation}

\noindent and after the substitution of $n = k^2 c^2 / \omega^2$, it becomes

\begin{equation}
	\frac{\omega^2}{k^2} \left( 1 + \frac{n_f}{n_s} + \frac{v_{\text{A}}^2}{c^2} \right) - 2 \frac{n_f}{n_s} v_0 \frac{\omega}{k} + \frac{n_f}{n_s} v_0^2 - v_{\text{A}}^2 = 0.
	\label{eq:awcfdi}
\end{equation}

\noindent This is the same dispersion relation as in Ref.~\cite{awcfdm}, except for the term $V_{\text{A}}^2 / c^2$ that here appears within the first bracket and is caused by the term $\omega^2 / k^2 {\bf E}$ of Eq.~(\ref{eq:afl}), which is neglected in~\cite{awcfdm} in the expression of the Amp\`ere law.

Equation (\ref{eq:rlinst}) therefore represents the generalization of the case where at low frequencies Alfv\'en wave couples with the flow driven ($\omega = k v_0$) mode~\cite{awcfdm}.  Eq.~(\ref{eq:rlinstii2}) can be solved analytically for ion/ion interactions. By the substitution of $R (L) = n_{\text R/L} = k^2 c^2 / \omega^2$ in Eq.~(\ref{eq:rlinstii2}), quartic polynomial equation is obtained:

\begin{eqnarray}
	&&a_i~\omega^4 + b_i~\omega^3 + c_i~\omega^2 + d_i~\omega + e_i = 0,
	\label{eq:wr4i}
	\\
	&&a_i = \frac{1}{\omega_{ci}^2} \frac{v_{\text{A}}^2}{c^2}, \nonumber
	\\
	&&b_i = \pm \frac{1}{\omega_{ci}} \left[ 1 + \eta + \frac{v_{\text{A}}^2}{c^2} \left( 2 \mp \frac{{\bf k}\cdot{\bf v}_0}{\omega_{ci}} \right) \right], \nonumber
	\\
	&&c_i = \left( 1 + \eta + \frac{v_{\text{A}}^2}{c^2} \right) \left( 1 \mp \frac{{\bf k}\cdot{\bf v}_0}{\omega_{ci}} \right) \mp \eta~\frac{{\bf k}\cdot{\bf v}_0}{\omega_{ci}} - \frac{k^2 v_{\text{A}}^2}{\omega_{ci}^2}, \nonumber
	\\
	&&d_i = \left( \eta~{\bf k}\cdot{\bf v}_0 \pm \frac{k^2 v_{\text{A}}^2}{\omega_{ci}} \right) \left( -2 \pm \frac{{\bf k}\cdot{\bf v}_0}{\omega_{ci}} \right), \nonumber
	\\
	&&e_i = \eta~({\bf k}\cdot{\bf v}_0)^2 - k^2 v_{\text{A}}^2 \left( 1 \mp \frac{{\bf k}\cdot{\bf v}_0}{\omega_{ci}} \right). \nonumber
\end{eqnarray}

\noindent General solutions to this equation are found as

\begin{eqnarray}
	&&\omega_{1,2} = - \frac{b_i}{4 a_i} - S_i \pm \frac{1}{2} \sqrt{- 4 S_i^2 - 2 p_i + \frac{q_i}{S_i}},
	\label{eq:w12i}
	\\
	&&\omega_{3,4} = - \frac{b_i}{4 a_i} + S_i \pm \frac{1}{2} \sqrt{- 4 S_i^2 - 2 p_i - \frac{q_i}{S_i}},
	\label{eq:w34i}
\end{eqnarray}

\noindent where coefficients and determinants are given by

\begin{eqnarray}
	&&p_i = \frac{8 a_i c_i - 3 b_i^2}{8 a_i^2}, \nonumber
	\\
	&&q_i = \frac{b_i^3 - 4 a_i b_i c_i + 8 a_i^2 d_i}{8 a_i^3}, \nonumber
	\\
	&&S_i = \frac{1}{2} \sqrt{ -\frac{2}{3} p_i + \frac{1}{3 a_i} \left( Q_i + \frac{\Delta_0}{Q_i} \right)}, \nonumber
	\\
	&&Q_i = \sqrt[3]{\frac{\Delta_1 + \sqrt{\Delta_1^2 - 4 \Delta_0^3}}{2}}, \nonumber
	\\
	&&\Delta_0 = c_i^2 - 3 b_i d_i + 12 a_i e_i, \nonumber
	\\
	&&\Delta_1 = 2 c_i^3 - 9 b_i c_i d_i + 27 b_i^2 e_i + 27 a_i d_i^2 - 72 a_i c_i e_i. \nonumber
\end{eqnarray}

\subsection{\label{sec:appparallelinstee}Electron-cyclotron frequencies}

When frequencies around the electron-cyclotron or higher ($\xi \omega \sim \omega_{ce} \gg \omega_{ci}$) are considered, and the condition in which the electron plasma frequency is much higher than the ion one ($\omega_{pe} \gg \omega_{pi}$) is assumed, Eq.~(\ref{eq:rlinst}) can be approximated by the expression:

\begin{equation}
	R (L) \approx 1 - \frac{\omega_{pes}^2}{\omega (\omega \pm\omega_{ce})} - \eta \frac{\xi^2 \omega_{pes}^2}{\xi \omega (\xi \omega \pm\omega_{ce})}.
	\label{eq:rlinstee}
\end{equation}

This quartic polynomial equation can be rewritten in the form similar to the case of the ion-cyclotron frequencies:

\begin{eqnarray}
	&&a_e~\omega^4 + b_e~\omega^3 + c_e~\omega^2 + d_e~\omega + e_e = 0, \label{eq:wr4e}
	\\
	&&a_e = \frac{1}{\omega_{ce}^2} \frac{v_{\text{A}e}^2}{c^2}, \nonumber
	\\
	&&b_e = \frac{1}{\omega_{ce}} \left[ - \frac{v_{\text{A}e}^2}{c^2} \left( \frac{{\bf k}\cdot{\bf v}_0}{\omega_{ce}} \mp 2 \right) \right], \nonumber
\end{eqnarray}
\begin{eqnarray}
&&c_e = \frac{v_{\text{A}e}^2}{c^2} \left( 1 \mp \frac{{\bf k}\cdot{\bf v}_0}{\omega_{ce}} \right) - \left(1 + \eta + \frac{k^2 v_{\text{A}e}^2}{\omega_{ce}^2} \right), \nonumber
	\\
	&&d_e = ( {\bf k}\cdot{\bf v}_0 \mp \omega_{ce} ) \left( 1 + \eta + \frac{k^2 v_{\text{A}e}^2}{\omega_{ce}^2} \right) \mp \frac{k^2 v_{\text{A}e}^2}{\omega_{ce}}, \nonumber
	\\
	&&e_e = \pm \eta~{\bf k}\cdot{\bf v}_0~\omega_{ce} - k^2 v_{\text{A}e}^2 \left( 1 \mp \frac{{\bf k}\cdot{\bf v}_0}{\omega_{ce}} \right). \nonumber
\end{eqnarray}

Solutions to this equation are found by the same pattern as given in Eqs.~(\ref{eq:w12i},~\ref{eq:w34i}) using the coefficients with the subscript $e$ instead of $i$. Contrary to the case of ions, instability due to the bulk flow of electrons does not arise here, and they are therefore not subjected to further analysis.

\section{\label{sec:appnormal}Instability of O and X waves}

We distinguish the two cases for the waves that propagate normal to ${\bf B}_0$. If the flow is parallel to the wavevector (${\bf v}_0 = v_0 {\bf \hat{z}}$), the polarization matrix given by Eq.~(\ref{eq:dm}) takes the form:

\begin{eqnarray}
	&&\mathcal{D} =
	\begin{pmatrix}
		P - n^2 & 0 & 0 \\
		0 & S - n^2 & -i D \\
		0 & i D & S \\
	\end{pmatrix} + \label{eq:dmnp}
	\\
	&&+ \sum_{\alpha = i,e} \frac{\omega_{p\alpha f}^2}{\omega^2} \cdot
	\begin{pmatrix}
			~~0~ & 0 & 0 \\
			~~0~ & 0 & - i \dfrac{\omega_{c\alpha} k v_0}{\xi^2 \omega^2 - \omega_{c\alpha}^2} \\
			~~0~ & i \dfrac{\omega_{c\alpha} k v_0}{\xi^2 \omega^2 - \omega_{c\alpha}^2} & -\dfrac{ \omega k v_0 }{\xi^2 \omega^2 - \omega_{c\alpha}^2}(\xi + 1) \\
	\end{pmatrix}. \nonumber
\end{eqnarray}

If the flow is perpendicular to the wavevector and parallel to ${\bf B}_0$ (${\bf v}_0 = v_0 {\bf \hat{x}}$), the scalar product ${\bf k} \cdot {\bf v}_0$ vanishes and $\xi = 1$, which reduces the Eq.~(\ref{eq:dm}) to

\begin{eqnarray}
	&&\mathcal{D} =
	\begin{pmatrix}
		P - n^2 & 0 & 0 \\
		0 & S - n^2 & -i D \\
		0 & i D & S \\
	\end{pmatrix} + \label{eq:dmnn}
	\\
	&&+ \sum_{\alpha = i,e} \frac{\omega_{p\alpha f}^2}{\omega^2} \cdot
	\begin{pmatrix}
			-\dfrac{k^2 v_0^2}{\omega^2 - \omega_{c\alpha}^2} & i \dfrac{\omega_{c\alpha} k v_0}{\omega^2 - \omega_{c\alpha}^2} & -\dfrac{\omega k v_0}{\omega^2 - \omega_{c\alpha}^2} \\
			- i \dfrac{\omega_{c\alpha} k v_0}{\omega^2 - \omega_{c\alpha}^2} & 0 & 0 \\
			-\dfrac{\omega k v_0}{\omega^2 - \omega_{c\alpha}^2} & 0 & 0 \\
	\end{pmatrix}. \nonumber
\end{eqnarray}

\subsection{\label{sec:appomode}Ordinary electromagnetic wave}

From previous relations, it can be seen that the O-mode is modified by the flow only if ${\bf v}_0 \parallel {\bf B}_0$, which is described by the first root of Eq.~(\ref{eq:dmnn}). For simplicity, we analyze this wave in the center-of-momentum reference frame. This means that we have the two opposite streaming plasmas, which satisfy the condition $\rho_1 v_{1} = -\rho_2 v_{2}$. Plasma frequencies and velocities are thus related as $\omega_{p2} = \eta \omega_{p1}$ and $v_{2} = -v_{1} / \eta$, respectively. The non-diagonal terms in Eq.~(\ref{eq:dmnn}) are therefore cancelled, and the first root simplifies:

\begin{equation}
	n^2 = P - \sum_{\alpha=i,e} \frac{\omega_{p\alpha 1}^2}{\omega^2} \frac{k^2 v_{1}^2}{\omega^2 - \omega_{c\alpha}^2} \left( 1 + \frac{1}{\eta} \right). \label{eq:dmom}
\end{equation}

If we consider unmagnetized plasmas ($\omega_{c\alpha} = 0$), Eq.~(\ref{eq:dmom}) reduces to a relation that has a purely complex solution. It represents the growth rate of Weibel instability~\cite{weibel}, which is commonly observed in shock simulations~\cite{unmagnetized,sda}. In the case of magnetized flows, if $\omega \sim \omega_{ci}$, Eq.~(\ref{eq:dmom}) simplifies to a biquadratic equation

\begin{eqnarray}
	\omega^4 &&- \omega^2 (\omega_p^2 + k^2 c^2 + \omega_{ci}^2) -
	\label{eq:drom}
	\\
	&&- \omega_{pi}^2 k^2 v_1^2 + \omega_{ci}^2 (\omega_p^2 + k^2 c^2) = 0, \nonumber
\end{eqnarray}

\noindent where $\omega_p^2 = \sum_{i,e} ( 1 + \eta ) \omega_{pi1}^2$ and $\omega_{pi} = ( 1 + 1 / \eta ) \omega_{pi1}$. The solution for the unstable wave mode, we find as

\begin{equation}
\omega^2 = - \omega_{pi}^2 \frac{v_1^2}{c^2} \left[ \frac{1}{1 + \dfrac{\omega_p^2}{k^2 c^2}} - \frac{1}{\left( 1 + \dfrac{1}{\eta} \right) M_\text{A}^2} \right].
\label{eq:w2om}
\end{equation}

\noindent For $B_0 = 0$, this solution necessarily becomes complex, and takes the known form of the growth rate of Weibel instability. However, in the case of magnetized plasmas, for $\omega \sim \omega_{ci}$, Eq.~(\ref{eq:w2om}) implies that the stability criterion now depends on the Alfv\'enic Mach number. For a given $M_\text{A}$, there is a cutoff wavenumber $k_{cut} \approx r_{gi}^{-1} \sqrt{\eta~m_i/m_e}$, at which the wave becomes unstable.

\subsection{\label{sec:appxmode}Extraordinary electromagnetic wave}

If the flow is along the magnetic field lines, the second root of Eq.~(\ref{eq:dmnn}), as given in the center-of-momentum frame, is

\begin{equation}
	(S - n^2) S - D^2 = 0,~~~n^2 = \frac{R L}{S}. \nonumber
\end{equation}

\noindent This mode is a simple elliptically polarized EM wave, with the electric field vector in a direction perpendicular to the magnetic field lines (${\bf E} \perp {\bf B}_0$).

If the flow is directed parallel (${\bf v}_0 = v_0 {\bf \hat{z}}$) to the wavevector, the X-mode becomes modified. Its dispersion relation is given by the second root of Eq.~(\ref{eq:dmnp}):

\begin{equation}
	(S - n^2) S_M - D_M^2 = 0,~n^2 = \frac{R L - S M_S - 2 D M_D - M_D^2}{S - M_S}, \nonumber
\end{equation}

\noindent where $S_M$ and $D_M$ are expressed as

\begin{eqnarray}
	S_M &&= S - \sum_{\alpha = i,e} \frac{\omega_{p\alpha f}^2}{\omega^2} \frac{ \omega k v_0 }{\xi^2 \omega^2 - \omega_{c\alpha}^2}(\xi + 1), \nonumber
	\\
	D_M &&= D + \sum_{\alpha = i,e} \frac{\omega_{p\alpha f}^2}{\omega^2} \frac{\omega_{c\alpha} k v_0}{\xi^2 \omega^2 - \omega_{c\alpha}^2}. \nonumber
\end{eqnarray}

\bibliographystyle{aipauth4-1}
\bibliography{rmiqpcs.bbl}

\end{document}